\documentclass[12pt]{spieman}  
\usepackage{amsmath,amsfonts,amssymb}
\usepackage{graphicx}
\usepackage{setspace}
\usepackage{tocloft}
\usepackage{macro_ref}

\graphicspath{ {./figures/} }
\usepackage{caption} 
\usepackage{subcaption}
\usepackage{lineno}
\usepackage{soul,color} 
\newcommand{\De}{\Delta}

\title{Achromatic design of a photonic tricoupler and phase shifter for broadband nulling interferometry}

\author[a,b,c]{Teresa Klinner-Teo}
\author[a,b,c,*]{Marc-Antoine Martinod}
\author[a,b,c]{Peter Tuthill}
\author[d]{Simon Gross}
\author[a,b,c]{Barnaby Norris}
\author[a,b,c]{Sergio Leon-Saval}
\affil[a]{Sydney Institute for Astronomy, School of Physics, University of Sydney, NSW 2006, Australia}
\affil[b]{Sydney Astrophotonic Instrumentation Laboratories, University of Sydney, NSW 2006, Australia}
\affil[c]{AAO-USyd, School of Physics, University of Sydney, NSW 2006, Australia}
\affil[d]{MQ Photonics Research Centre, School of Engineering, Macquarie University, NSW 2006, Australia}

\cftpagenumbersoff{figure}
\cftpagenumbersoff{table} 
\begin{document} 
\maketitle

\begin{abstract}
    Nulling interferometry is one of the most promising technologies for imaging exoplanets within stellar habitable zones. 
    The use of photonics for carrying out nulling interferometry enables the contrast and separation required for exoplanet detection.
    So far, two key issues limiting current-generation photonic nullers have been identified: phase variations and chromaticity within the beam combiner.
    The use of tricouplers addresses both limitations, delivering a broadband, achromatic null together with phase measurements for fringe tracking.
    Here, we present a derivation of the transfer matrix of the tricoupler, including its chromatic behaviour, and our 3D design of a fully symmetric tricoupler, built upon a previous design proposed for the GLINT instrument. It enables a broadband null with symmetric, baseline-phase-dependent splitting into a pair of bright channels when inputs are in anti-phase.
    Within some design trade space, either the science signal or the fringe tracking ability can be prioritised. 
    We also present a tapered-waveguide $180^\circ$-phase shifter with a phase variation of $0.6^\circ$ in the $1.4-1.7~\mu$m band, producing a near-achromatic differential phase between beams{ for optimal operation of the tricoupler nulling stage}.
    Both devices can be integrated to deliver a deep, broadband null together with a real-time fringe phase metrology signal. 
\end{abstract}


{\noindent \footnotesize\textbf{*}Email:  \linkable{marc-antoine.martinod@sydney.edu.au} }

\begin{spacing}{1}   

\section{Introduction} \label{sect:intro}
    
    Methods for exoplanet direct detection and characterisation are often limited by the inherent high contrast ratio and small angular separation intrinsic to exoplanetary systems, even when found orbiting the most favorable nearby stars. 
    Unlike indirect detection methods, nulling interferometry enables direct imaging of dim stellar companions, extincting the bright starlight by arranging a condition of destructive interference from two (or more) apertures \cite{bracewell1978detecting}. 
    Interferometry can also exploit sparse or separated apertures and is not bound by the formal diffraction limit of conventional monolithic telescopes, and so can observe companions closer to the host star compared with techniques such as coronagraphy\cite{norris2020first}. 
    The contrast ratios required for exoplanet detection span a range from $10^{-4}$ for self-luminous hot exoplanets observed in the mid-infrared \cite{Marois2008} to $10^{-10}$ for Earth-like exoplanets imaged in reflected light from their host star \cite{Schworer2015}.
    Performance levels over this range lie within theoretical limits for nulling interferometers (``nullers'' hereafter). 
    While most have hitherto been built using bulk optics (for example, the Keck interferometer\cite{colavita2009keck} and the Large Binocular Telescope Interferometer\cite{defrere2016nulling}), integrated optics and photonic methods are increasingly becoming the desired technology for nullers\cite{mennesson2011high}. 
    {The use of photonics provides single-mode waveguides, which perform modal-filtering (where phase aberrations from the wavefront translate coupling efficiencies into single-moded waveguides, in essence converting phase into intensity fluctuations), and their monolithic designs ensure stability against environmental conditions and compactness for scalability\mbox{\cite{norris2020first}}}. 
    
    {The Guided-Light Interferometric Nulling Technology (GLINT) instrument\mbox{\cite{norris2020first, martinod2021scalable}} is a nuller currently integrated into the Subaru Coronagraphic Extreme Adaptive Optics (SCExAO) system at the Subaru Telescope, and is operated by our team}. It is able to combine light in the astronomical $H$ band from multiple sub-apertures coupled into separate single-mode waveguides, nulling multiple baselines simultaneously.
    In its present configuration, nulling is carried out within a photonic chip, which is able to combine light channeled from four sub-apertures on the telescope, producing six non-redundant baselines. 
    Light is combined using directional couplers, with output flux directed into two channels: (1) the \textit{null} channel in which on-axis starlight destructively interferes and is therefore heavily suppressed (potentially, together with off-axis planet-light not similarly extincted), and (2) the \textit{anti-null} or ``bright'' channel which contains constructively interfered starlight. 
    The aim of the instrument is to tune the interference to maximise destructive interference of the starlight, extincting it from the null channel and routing it instead into the bright channel. 
    The most commonly used observable quantity, the null depth, is derived from the ratio of these two outputs. 
    The GLINT device also features photometric monitoring outputs that give an instantaneous measurement of the flux contributions of each input beam to the interference signals.
    This configuration of the instrument has been characterised on-sky and demonstrated\cite{martinod2021scalable} to reach a null depth of $10^{-3}$ with a precision of $10^{-4}$.
    However, to be truly scientifically competitive, a nuller targeting planets in the infrared\cite{serabyn2000nulling} will require a significantly improved null depth of around $10^{-6}$.
    The current limitations preventing GLINT from reaching deeper nulls have been identified: the chromatic behavior of the directional coupler that induces a wavelength-dependent null depth and the residual phase fluctuations of light entering the instrument.
    Replacing the directional coupler with a symmetric tricoupler, which splits input beams into three outputs, rather than two, has been shown in simulation to be capable of overcoming these limitations\cite{martinod2021achromatic}.
    Assuming that the waveguides in the interaction region of the tricoupler are arrayed in an equilateral triangle and that the injected beams are in perfect anti-phase and with equal flux, the central channel will produce a completely nulled output, regardless of the wavelength of incoming light. 
    Underlying this powerful configuration are arguments based on simple symmetry: anti-symmetric input beams are unable to couple to an even-symmetry mode field required to overlap with the central null channel\cite{labeye2004}.
    A further advantage is provided by the output fluxes from the two bright channels, which can be used to measure the incoming differential phase thereby providing a signal for active cophasing correction in real time.
    Simulations of a tricoupler able to deliver this achromatic null have been shown to reach a null depth 45 times better than with a single directional coupler \cite{martinod2021achromatic}.
    The use of a tricoupler has been previously mooted as a combiner for nulling and stellar interferometry \cite{weber2004, labeye2004}. 
    {To exploit the equal coupling ratios and phase delays between outputs, early devices made use of three waveguides in a planar, rather than triangular, configuration. 
    In order to produce the desired performance, waveguides were tapered to adiabatically alter the coupling ratios\mbox{\cite{vance1994design}}. Planar tricouplers following this methodology have been designed for nulling interferometry\mbox{\cite{hsiao2010}}. 
    However, devices with an equilateral triangular configuration in the interaction region are able to exploit further symmetries: namely, an equal splitting ratio between bright channels\mbox{\cite{martinod2021achromatic}}.
    Early tricouplers made use of fused fibre twisted together to produce this triangular configuration in the central region, which introduced asymmetries, altering the desired coupling ratio \mbox{\cite{birks1992effect}}. 
    Today, the ultrafast laser inscription (ULI)\mbox{\cite{nolte2003femtosecond,gattass2008femtosecond,arriola2013low,gross2015ultrafast}} process enables efficient and precise inscription of tricouplers and other photonic devices requiring three-dimensional design freedom. 
    The use of tricouplers for space interferometry is currently being investigated\mbox{\cite{Hansen2020, hansen2022interferometric}}.}
    
    Here, we present a full chromatic treatment of a new design for a tricoupler. 
    This design incorporates fully symmetric triangular inputs and outputs, unlike the design presented in Ref.~\citenum{martinod2021achromatic}, which eliminates slight asymmetries in the splitting ratios and allows for equal splitting between non-null channels across the band. 
    This greatly simplifies the treatment of interactions using the tricoupler's transfer matrix and offers more flexibility with choice of waveguide injected into, while still being fully fabricable using the ULI technique. 
    Additionally, we investigate the potential for this tricoupler to maximise the signal from a faint stellar companion, showing that adjustments to the physical properties of the device can either maximise science throughput or provide better fringe tracking. 
    Furthermore, any instrument incorporating a tricoupler for the nulling beam combination requires a perfect $180^\circ$ differential phase between injected beams across the working spectral band.
    Therefore, an achromatic phase shift has to be implemented preceding every tricoupler in the chip to enforce this anti-phase condition between beams.
    To date, simulations of nulling with tricouplers have assumed inputs arranged to be in anti-phase using a (chromatic) delay in air: a simple piston term applied to one beam\cite{martinod2021achromatic} which only meets the ideal anti-phase condition at one wavelength within the observing band.
    This limits the maximum number of simultaneously nulled baselines for any given number of apertures and wastes the signal-to-noise advantages that a truly achromatic null would deliver over all spectral channels in the band.
    We introduce an achromatic $180^\circ$ photonic phase shifter using tapered waveguides, which can be used in concert with the tricoupler to achieve an achromatic null across the working band. 

    In Section~\ref{sec:modelling}, we derive the full transfer matrix for the current proposed tricoupler, taking chromatic behaviour into account. 
    We introduce our new proposed completely symmetric equilateral tricoupler in Section~\ref{sec:designs_tricoupler}, and simulate the throughput of a planet light signal, as well as examine the effects of changing the design to maximise this signal, in Section~\ref{sec:planetlight}. 
    In Section~\ref{sec:designs_aps}, we advance our design for an achromatic phase shifter which can be added at the input side of each tricoupler. The combination of both devices can be replicated for each combiner in the case of a multi-baseline nuller, as shown in Section~\ref{sec:device}.
    These photonic components can be realistically fabricated using ULI inside a single integrated-optics chip.

\section{Modelling beam combination with transfer matrices}\label{sec:modelling}

\begin{figure}[h]
        \centering
        \includegraphics[width=0.8\textwidth]{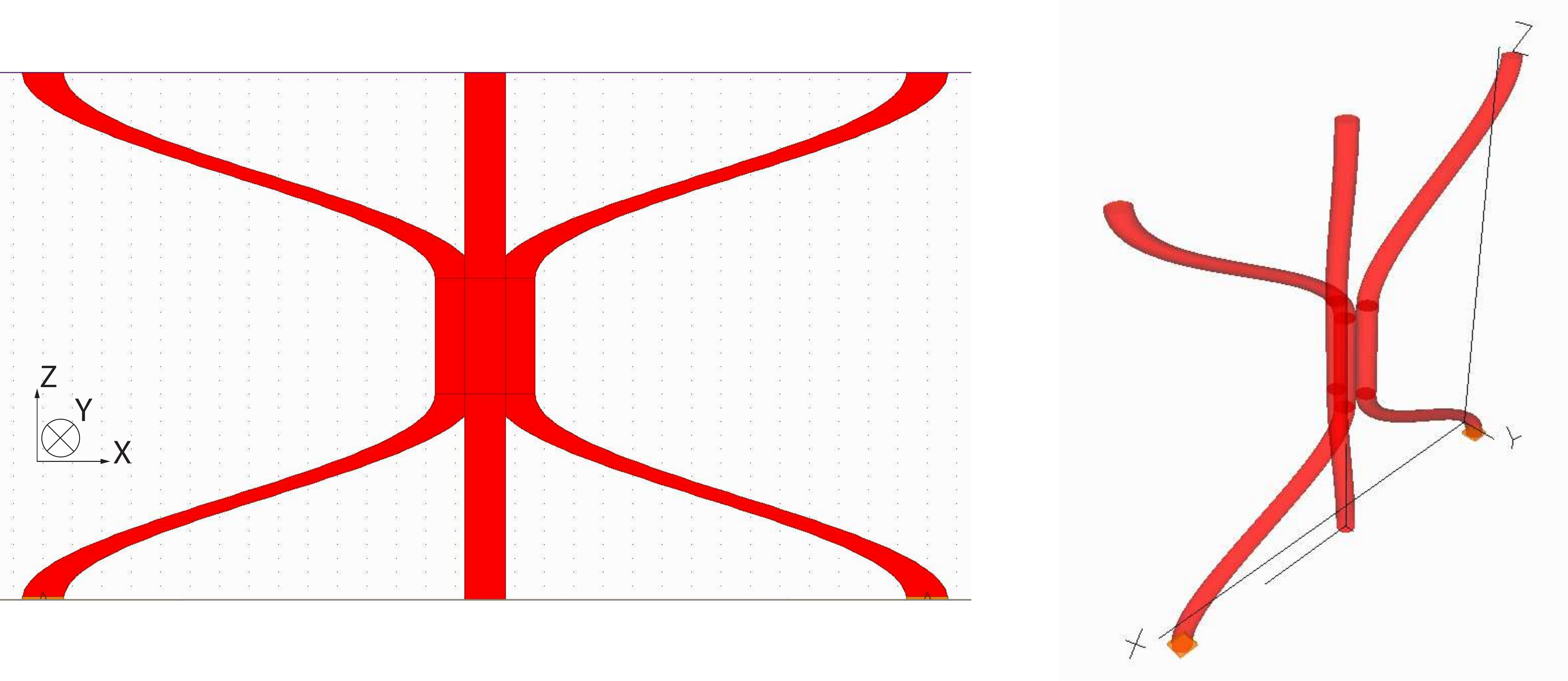}
        \caption{{Top (left) and 3D (right) view of the asymmetric tricoupler. While the interaction region follows an equilateral triangular pattern, some coupling effects happen in the remapping from and to a linear array before and after the interaction region, breaking the full symmetry of the interactions between the waveguides inside the tricoupler. The light enters the device from the bottom and leaves at the top of the tricoupler.}}
        \label{fig:asym_tricoupler_design}
    \end{figure}

    The tricoupler proposed in Ref.~\citenum{martinod2021achromatic} exhibits asymmetric coupling {ratios}.
    {While we expect similar coupling effects regardless of the waveguide in which the light is injected, some discrepancies arise depending on which waveguide is fed.
    These come from the waveguides' shape, which transitions from a linear array at both the inputs and outputs to an equilateral triangular configuration in the central region}. 
    {These transition areas result in two asymmetrically arranged zones at both ends of the interaction region, where the waveguides converge on approach to the interaction region, and again where they diverge \mbox{(Fig.\,\ref{fig:asym_tricoupler_design})}}.
    
\begin{figure}[h]
        \centering
        \includegraphics[width=0.8\textwidth]{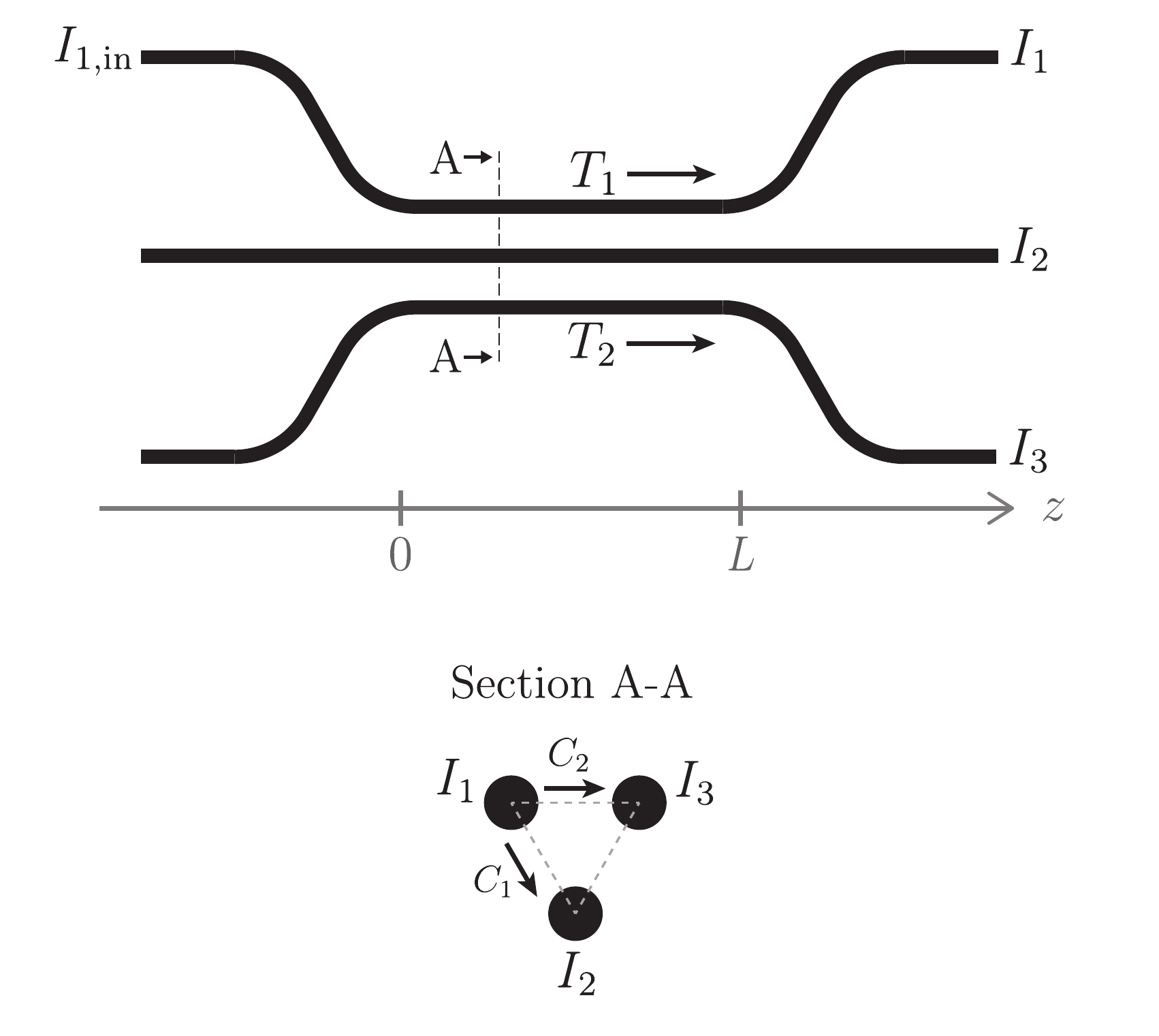}
        \caption{{Diagram of a tricoupler. Thick black lines represent waveguides. Section A-A shows a cross-section of the interaction region in the $xy$-plane, where the three waveguides are arranged in an equilateral triangular formation. The interaction region has length $L$. The top waveguide is illuminated with intensity $I_\text{1,in}$, and couples into the other waveguides with coupling coefficients $C_1$ and $C_2$, and transmission coefficients $T_1$ and $T_2$. }}
        \label{fig:tricoupler_coeffs_diagram}
    \end{figure}

    The transfer matrix for a tricoupler with symmetry between left and right waveguides, but not three-fold equilateral symmetry as described above, would have at least two independent coupling coefficients. Hence mode field orthogonality is generally needed, in addition to energy conservation and symmetry arguments, to determine the phase shift. However, by assuming mode orthogonality principle\cite{fang1996interferometric}, the transfer matrix can be written as
    \begin{equation}\label{asym_matrix_phasor}
        M = 
        \begin{pmatrix}
            \Tilde{T}_1 & \Tilde{C}_1 & \Tilde{C}_2 \\
            \Tilde{C}_1 & \Tilde{T}_2 & \Tilde{C}_1 \\
            \Tilde{C}_2 & \Tilde{C}_1 & \Tilde{T}_1 \\
        \end{pmatrix},
    \end{equation}
    where $\Tilde{C}_1$, $\Tilde{C}_2$,  $\Tilde{T}_1$ and $\Tilde{T}_2$ are the complex coupling coefficients from the input waveguide to the central and outer output waveguides, and transmission coefficients in the outer and central waveguides, respectively {\mbox{(Fig.\,\ref{fig:tricoupler_coeffs_diagram})}}.
    
    $M$ describes the relationship between the input and output beams of the tricoupler, such that if $I$ is a vector of the complex amplitudes of the input waveguides, then the vector of output waveguide amplitudes $O$ is given by $O=MI$. 
    The transmission and coupling coefficients are treated as complex phasors which act on the incoming beams $\Tilde{E}_{\text{in}}$:
    \begin{align}
        \Tilde{T}_i = T_ie^{i\psi_{T_i}} \\
        \Tilde{C}_i  = C_ie^{i\psi_{C_i}}
    \end{align}
    
     $T_i,\ C_i$ are defined to be the square modulus of these phasors such that
    
    \begin{equation}
        |\Tilde{T_i}|^2 = \frac{|\Tilde{E}_{\text{{out,i}}}|^2}{|\Tilde{E}_{\text{{in,i}}}|^2} = T_i^2.
    \end{equation}
    
    A tricoupler with asymmetric coupling or tapered waveguides\cite{hsiao2010,fang1996interferometric} requires the treatment of coupling to the centre and unfed outer waveguides, $C_1$ and $C_2$, separately. 
    As a result, expressions for the differential phases between channels become complicated to compute in terms of the coupling and transmission coefficients, which in turn makes the transfer matrix complicated and potentially impacts the efficiency of phase calculations for fringe tracking. 
    Instead, we focus here on a design of a fully symmetric tricoupler.
    Equal coupling of light simplifies the transfer matrix to
    \begin{align}
        M = 
        &\begin{pmatrix}
            \Tilde{T} & \Tilde{C} & \Tilde{C} \\
            \Tilde{C} & \Tilde{T} & \Tilde{C} \\
            \Tilde{C} & \Tilde{C} & \Tilde{T} \\
        \end{pmatrix} \nonumber\\
        = e^{i\psi_c}
        &\begin{pmatrix}\label{eqn:sym_tricoupler_matrix}
            Te^{i\Delta\psi} & C & C \\
            C & Te^{i\Delta\psi} & C \\
            C & C & Te^{i\Delta\psi} \\
        \end{pmatrix},
    \end{align}
    
    where, as above, the transmission and coupling coefficients $T$ and $C$ are wavelength-dependent and experimentally measurable by injecting light into one waveguide at the time, $\psi_t$ and $\psi_c$  are the arguments of the transmission and coupling coefficients respectively, and $\Delta\psi = \psi_t - \psi_c$. 
    
    {The conservation of energy between the inputs and the outputs implies\mbox{\cite{fang1996interferometric}}}
    \begin{equation}\label{phasor_reln}
        T^2 + 2C^2 = 1.
    \end{equation}
    {Hence the transmission coefficient is expressed with respect to the coupling coefficient as:}
    \begin{equation}
        \label{phasor_reln2}
        T = \sqrt{1 - 2C^2}.
    \end{equation}
    {Furthermore, the conservation of energy inside the device\mbox{\cite{fang1996interferometric}} itself implies that}
    \begin{align}
        |\Tilde{T} + 2\Tilde{C}|^2 = 1.
    \end{align}
    {Developing this and injecting the result of Equation \mbox{\ref{phasor_reln2}} gives}
    \begin{equation}\label{eqn:phase}
        \Delta\psi = \arccos{\frac{-C}{2T}}.
    \end{equation}
    Thus, the change of phase between waveguides can be characterised entirely by the measurement of the transmission and coupling coefficients. 
    
    {Interferometric signals collected in the three outputs can be obtained by considering two wavefronts $\tilde{a_1}$ and $\tilde{a_2}$ so that}
    \begin{align}
        \label{eq:interf_tricoupler}
         O = T_{\text{tricoupler}}
        &\begin{pmatrix}
            a_1 e^{i \phi_1}\\
            0 \\
            a_2 e^{i \phi_2}\\
        \end{pmatrix},
    \end{align}
    {where $O$ is the vector of the three output wavefronts, $a_i$ and $\phi_i$ are the amplitude and the phase of the incoming wavefront $i=1,2$.
    The intensities of the interferometric signals from the three outputs $I_A$, $I_B$ and $I_C$ are deduced from the development of the squared modulus of \mbox{Eq.~\eqref{eq:interf_tricoupler}} so that}
    \begin{align}
        \label{eq:interf_tricoupler2}
         \begin{pmatrix}
            I_A\\
            I_B \\
            I_B\\
        \end{pmatrix}
        =
        &\begin{pmatrix}
            T^2 I_1 + C^2 I_2 + 2 TC \sqrt{I_1 I_2} \cos \left(\phi_1 - \phi_2 + \Delta \psi \right)\\
            C^2 \left( I_1 + I_2 + 2 \sqrt{I_1 I_2} \cos \left(\phi_1 - \phi_2 \right) \right)\\
            C^2 I_1 + T^2 I_2 + 2 TC \sqrt{I_1 I_2} \cos \left(\phi_1 - \phi_2 - \Delta \psi \right)\\
        \end{pmatrix},
    \end{align}
   { where $I_i = a_i^2$ ($i=1,2$).
   We can see that the tricoupler indeed provides a nulled signal in the central output (second line of \mbox{Eq.~\eqref{eq:interf_tricoupler2}}) regardless the wavelength, providing $\phi_1 - \phi_2 = \pi$ and $I_1 = I_2$.}
    
    Designing the tricoupler so that $C=T=\frac{1}{\sqrt{3}}$ leads to a phase-shift $\Delta \psi = \frac{2\pi}{3}$ between all waveguides.
    This configuration is ideal for simultaneous broadband null depth measurement and fringe tracking capability\cite{martinod2021achromatic}.
    While careful design of 3-way planar waveguides has been shown capable of approaching this ideal\cite{hsiao2010}, doing so requires considerable control over many parameters governing the detailed waveguide properties.
    Here, instead, we adopt the strategy of enforcing equal coupling by way of geometrical symmetry.

\section{Designs for a symmetric equilateral tricoupler}\label{sec:designs_tricoupler}

    {In this Section and Section \mbox{\ref{sec:planetlight}}, we present two tricouplers, Tricoupler A and Tricoupler B, which are built upon previous designs and investigate the effects of tuning various parameters to optimise for equal splitting, fringe tracking or science signal performance. }

\begin{figure}[h]
        \centering
        \includegraphics[width=0.9\textwidth]{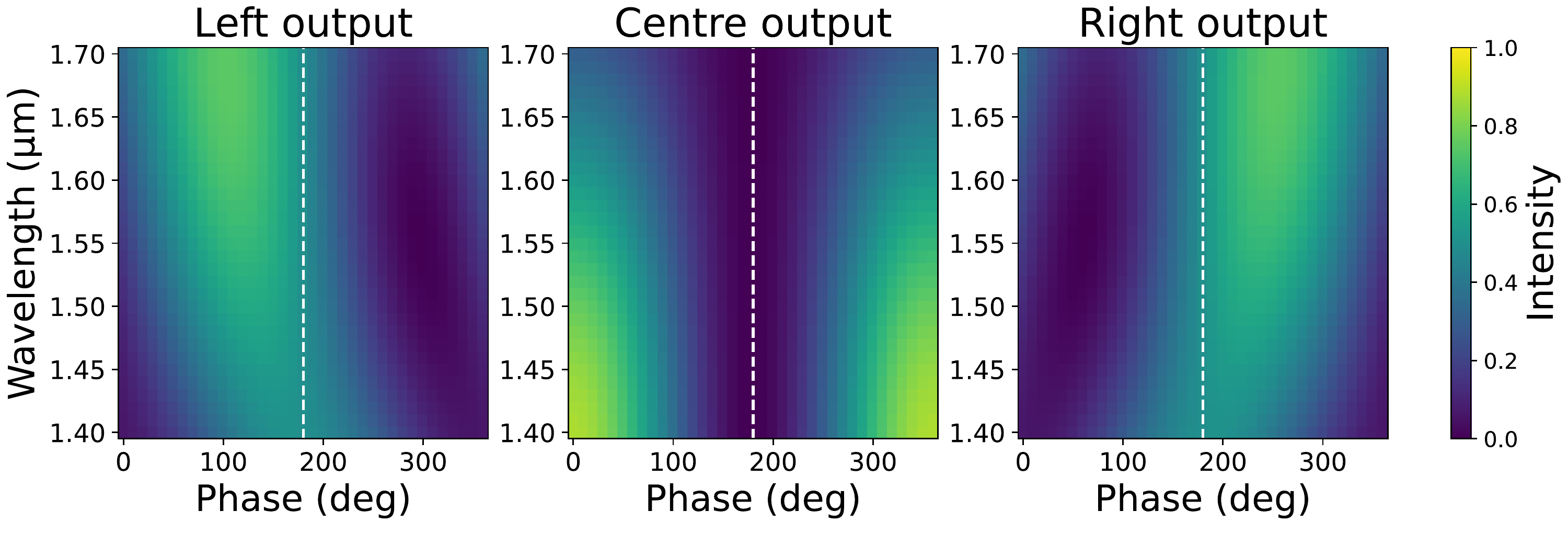}
        \caption{Map of the intensities in the three outputs of the tricoupler tuned for equal splitting at 1.55$~\mu$m, against wavelength and differential phase of incoming beams, when light of equal amplitude and phase is injected into the two input waveguides. Flux at 180$^\circ$ differential phase is shown by the white dashed line. Total intensity is normalised to 1.}
        \label{fig:tric_a_cmap}
    \end{figure}

    {As mentioned above, the tricoupler contains several exploitable symmetries when arranged in an equilateral triangular configuration. 
    The ideal tricoupler couples light with equal splitting ratios between all three waveguides, and outputs have a phase delay of $120^\circ$ between them. 
    Figure \mbox{\ref{fig:tric_a_cmap}} shows the output intensity of each channel as a function of wavelength and input differential phase for Tricoupler A, introduced in this section.}
    
    {When inputs have a differential phase of $180^\circ$, the central channel remains dark across the band.
    Furthermore, the left and right outputs are symmetric, meaning that the input differential phase -- even when not in perfect antiphase -- can be recovered.}

\begin{figure}[h]
        \centering
        \includegraphics[width=0.9\textwidth]{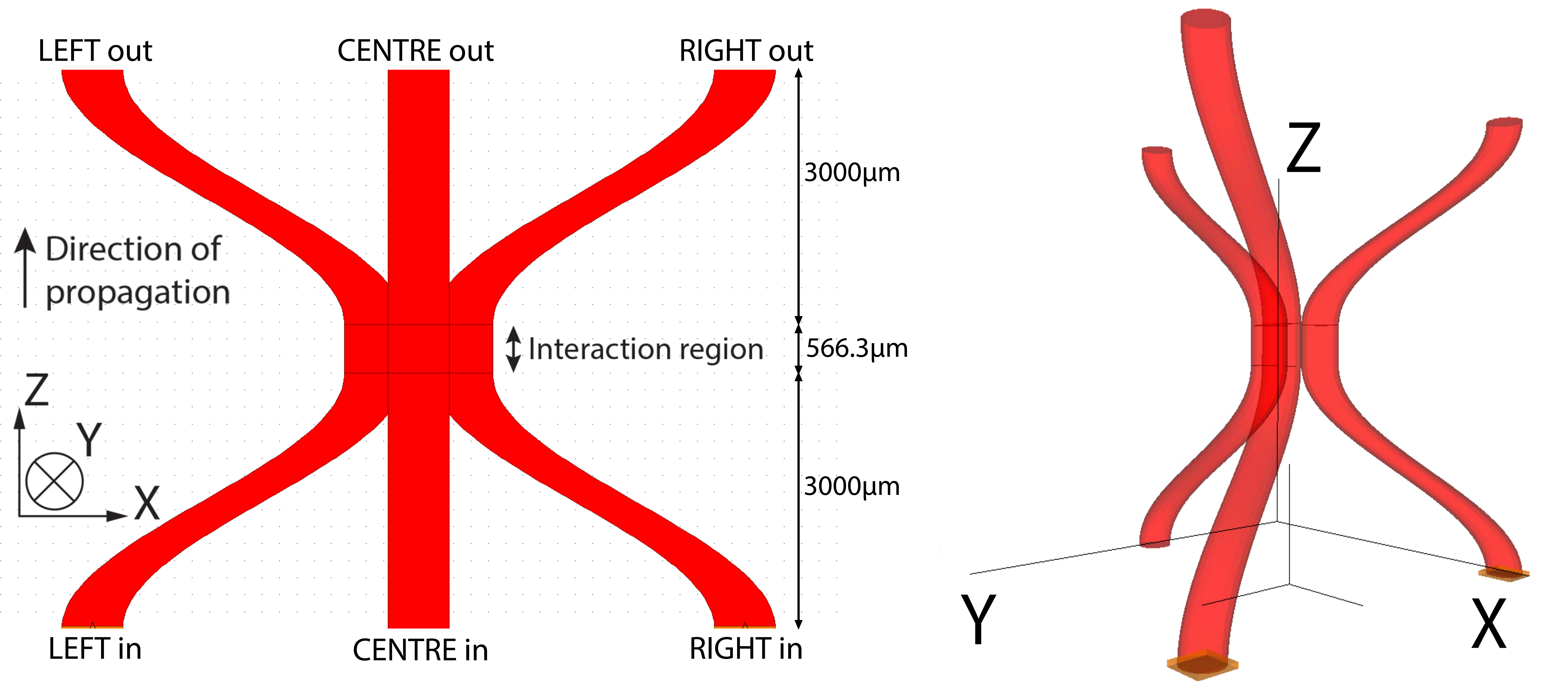}
        \caption{Schematic of the proposed equilateral tricoupler. The three inputs are at the bottom and direction of propagation is upward. Left: top view of the design. Right: 3D view of the design. Axes are not to scale. The separation between waveguides at input and output is 75$~\mu$m.}
        \label{fig:equil tricoupler}
    \end{figure}

    Our proposed design for a completely equilateral tricoupler{, Tricoupler A,} was created in the \texttt{RSoft} numerical simulation environment and is shown in Figure \ref{fig:equil tricoupler}.
    Unlike the device in Ref.~\citenum{martinod2021achromatic}, the branches of this tricoupler, as well as the central interaction region, are all arrayed in an equilateral triangular configuration. 
    The input beams are fed into two of the tricoupler's arms (the bottom left and right in Figure \ref{fig:equil tricoupler}), interfere with each other in the interaction region in the centre and are output at the top. The interaction region of{ Tricoupler A} has a length of $566.3~\mu$m, which allows for an equal 3-way ($\frac{1}{3}/\frac{1}{3}/\frac{1}{3}$) splitting ratio at the central wavelength of $1.55~\mu$m. Waveguides are labelled using the same convention as in Ref.~\citenum{martinod2021achromatic}. As in the previous design, the index difference between core and cladding is $0.005$, and the waveguide cross-section is circular, with a diameter of $7~\mu$m. 
    
\begin{figure}[h]
        \centering
        \begin{subfigure}[t]{0.65\textwidth}
          \includegraphics[width=1\textwidth]{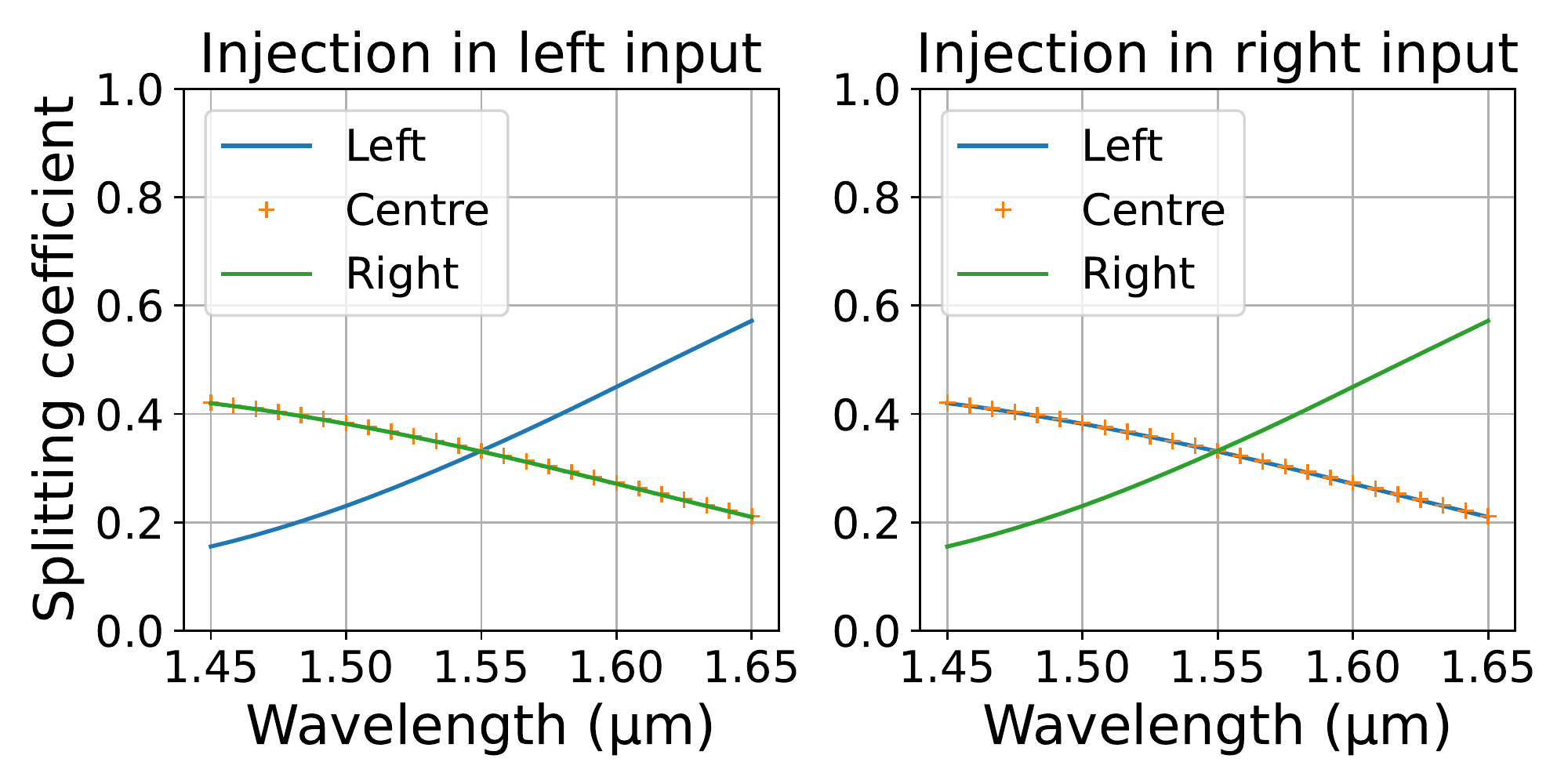}
          \caption{Equilateral symmetry tricoupler}
          \label{fig:equil_splittingcoeffs} 
        \end{subfigure}
        ~
        \begin{subfigure}[t]{0.65\textwidth}
          \includegraphics[width=1\textwidth]{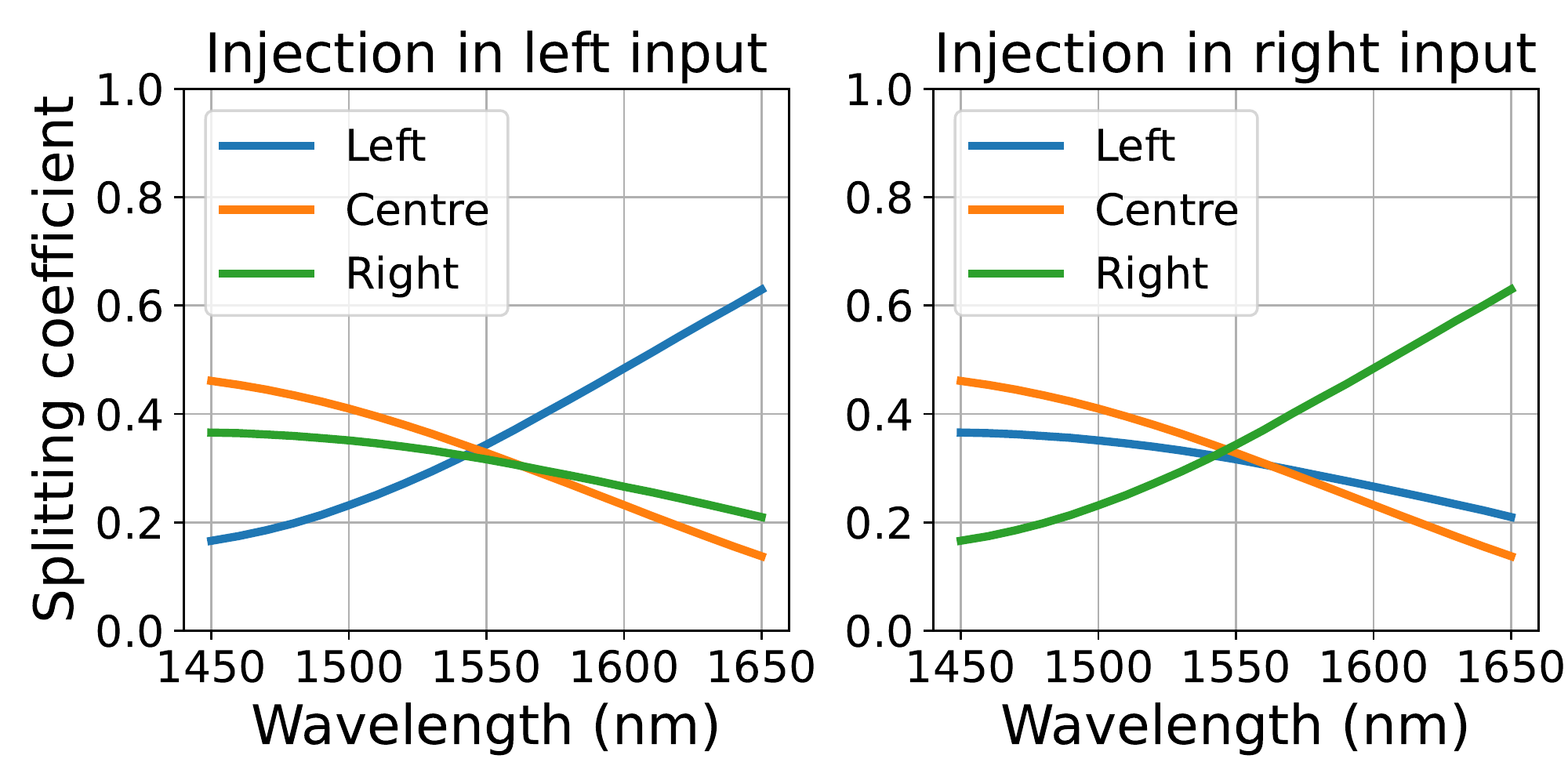}
          \caption{Left-right symmetry tricoupler\cite{martinod2021achromatic}}
          \label{fig:lr_splittingcoeffs}
        \end{subfigure}
        
        \caption{Intensity splitting ratios between the three outputs for injection of light into the left (left) and right (right) waveguide. The sum of the intensities is normalised to $1$ for each wavelength. Top: splitting ratios for the tricoupler presented here. Bottom: splitting ratios for the tricoupler presented in Ref.~\citenum{martinod2021achromatic}.}
        \label{fig:tricoupler_splittingcoeffs}
    \end{figure}

    In common with the earlier-generation tricoupler (symmetry only in the left and right waveguides\cite{martinod2021achromatic}), weak interaction between waveguides still occurs outside the interaction region, however the coupling ratios now remain equal (Fig.~\ref{fig:equil_splittingcoeffs}) due to the equilateral configuration, unlike the previous design which only manifests reflection symmetry (Fig.~\ref{fig:lr_splittingcoeffs}).
    Light injected at the central wavelength in the new design therefore splits equally between the two remaining waveguides. 
    This equal splitting allows interactions in the tricoupler to be described by the transfer matrix shown in Equation~\ref{eqn:sym_tricoupler_matrix}, which greatly simplifies calculations for both simulations of the chip and phase recovery for fringe tracking. 
    The design additionally exhibits greater flexibility in the choice of waveguide injection, as the equilateral symmetry allows for any of the three waveguides to be treated as the science channel, with the remaining two used as the bright channels for fringe tracking. 
    
    The change in tricoupler design leading to this additional symmetry comes at little cost to the convenience of its manufacture and integration into the nulling chip. 
    Although this design sacrifices the convenience of the planar configuration of input and output beams, its integration into the photonic circuitry can easily be accommodated. 
    At input, only two waveguides are illuminated before the tricoupler, so no significant symmetry-enforced changes are needed at the pupil remapper side with respect to the existing chip.
    At the output side, a modest extension of the triangular waveguides was found to be sufficient to separate them beyond any residual parasitic interaction.
    After this, reconfiguring them into a linear array (now without the constraint of optical pathlength matching) can be trivially accomplished.
    The ultrafast laser inscription technique, allowing the creation of 3-dimensional circuitry of almost arbitrary configuration, can easily meet all these practical design requirements.
    
\section{Considerations for science signal throughput}\label{sec:planetlight}

\begin{figure}[h]
        \centering
        \includegraphics[width=0.7\textwidth]{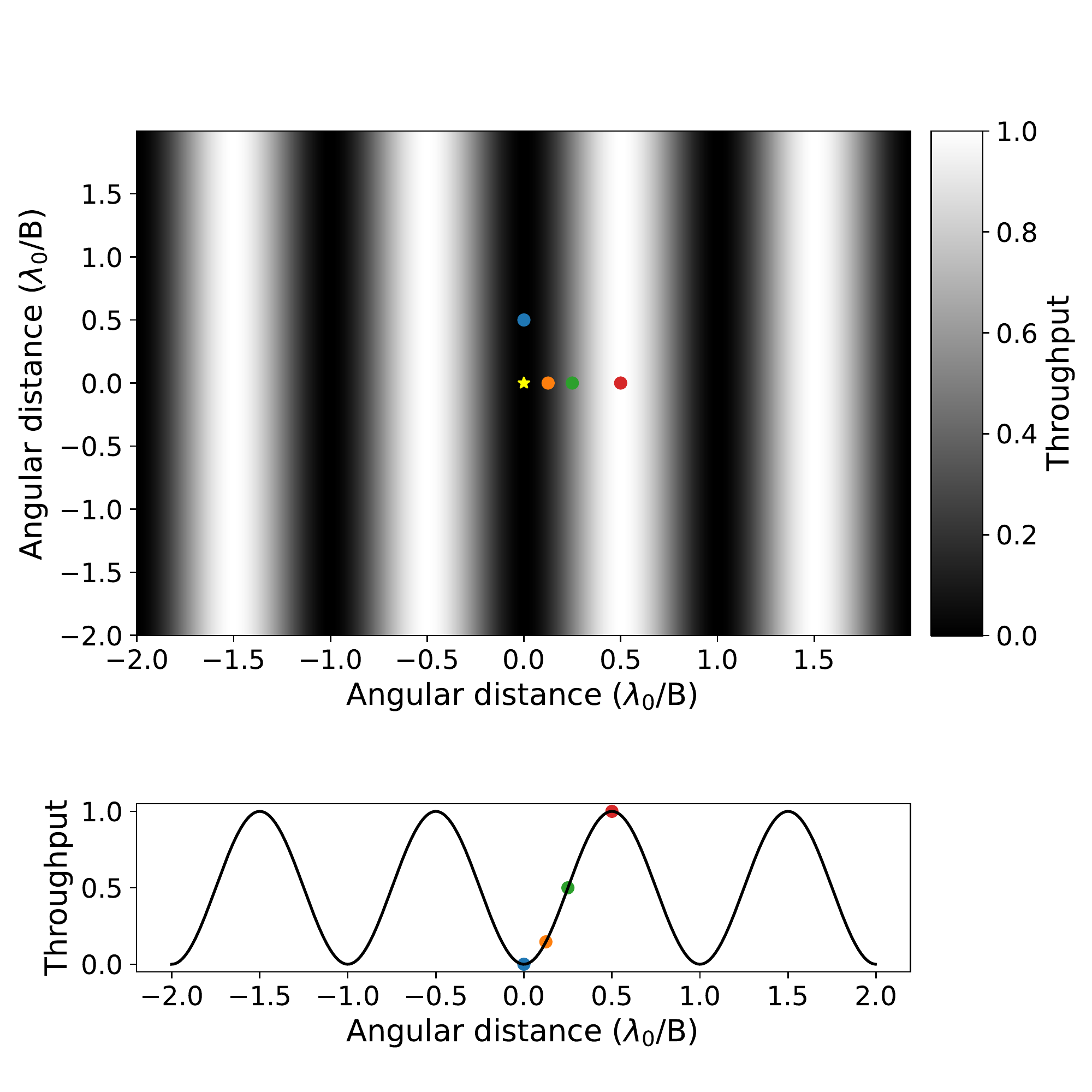}
        \caption{{Diagram of the fringe pattern created by a monochromatic nuller with candidate locations for the target star and planets overplotted. The yellow star denotes the targeted host star; the blue dot is a planet located in the central dark fringe of the nuller, the orange one is a planet located an eighth of a fringe away from the star, the green dot is a planet located a quarter of fringe away and the red dot is a planet in the centre of a bright fringe (half a fringe away from the star), which is the ideal situation for detecting its light. The bottom panel shows the ideal throughput of each planet as given by the nuller's fringe pattern.}}
        \label{fig:null_modulation}
    \end{figure}

    In addition to the ability to carry out achromatic nulling of starlight and fringe tracking, the capability of the tricoupler to transmit off-axis light from potential stellar companions (for convenience hereafter denoted ``planet light'') must also be considered. 
    {The nuller can be conceived as broadcasting a fringe pattern on-sky representing the transmission, with the central, dark fringe on the axis where it extincts flux from the target host star (Figure \mbox{\ref{fig:null_modulation}}). }
    In an ideal case, an exoplanet will be located in the bright fringe of the nuller, and therefore its light arrives at the nulling instrument with a differential phase of 180$^\circ$.

    {As this phase shift is due to a geometrical delay, there is a wavelength dependence in the differential phase across the band for the science object.
    On top of this astrometric effect, the tricoupler exhibits a chromatic throughput when the incoming phase is no longer equal to 180$^\circ$ (\mbox{Fig.~\ref{fig:tric_a_cmap}}).
    Therefore, the spectrum of the off-axis source is expected to be changed by the off-axis chromatic response of the tricoupler.
    An ideal tricoupler would exclude all incoming on-axis starlight while transmitting all incoming (off-axis) planet light in the central null output channel. 
    It is highly desirable to have an off-axis throughput as achromatic as possible to ensure sufficient signal to noise ratio at all wavelengths, as desired for spectroscopic measurements on the object.
    This property is of little consequence to the fringe tracking and phase stabilisation of the tricoupler, as the contribution of the planet light is negligible when compared with the starlight, due to the high-contrast nature of these observations. }
    
    The signal of an exoplanet located in the first bright fringe of the nuller's throughput was simulated using \texttt{RSoft}. 
    The angular offset $\theta$ of the planet from its host star was first calculated using the condition that the planet occurs in a bright fringe of the nuller, and has a differential phase of 180$^\circ$:
    \begin{align}
        \theta = \frac{\lambda_0}{2B},
    \end{align}
    where $\lambda_0$ is the wavelength in the centre of the band, $1.55~\mu$m, and $B$ is the baseline, which is set to 5.55m\cite{norris2020first}. 
    From this, the planet's wavelength-dependent differential phase at input was found:
    \begin{align}
        \De\psi = \frac{2\pi}{\lambda} B\theta,
    \end{align}
    where $\lambda$ is the wavelength of the incoming planet light. For this simulation, wavelengths in the $1.4 - 1.7~\mu$m range were spanned. 
    This phase offset planet signal was then fed into three different candidate tricouplers in order to explore the trade space driving the design process. 
    {Two key parameters were tuned: the pitch (separation between waveguides in the interaction region) and interaction length (length of the interaction region). As the starlight cancellation property of the tricoupler is affected only by the symmetry of the interaction region, by retaining the equilateral triangular configuration of all the tested devices, the broadband nulling capability was preserved. }
    
\begin{figure}[h]
        \centering
        \includegraphics[width=0.9\textwidth]{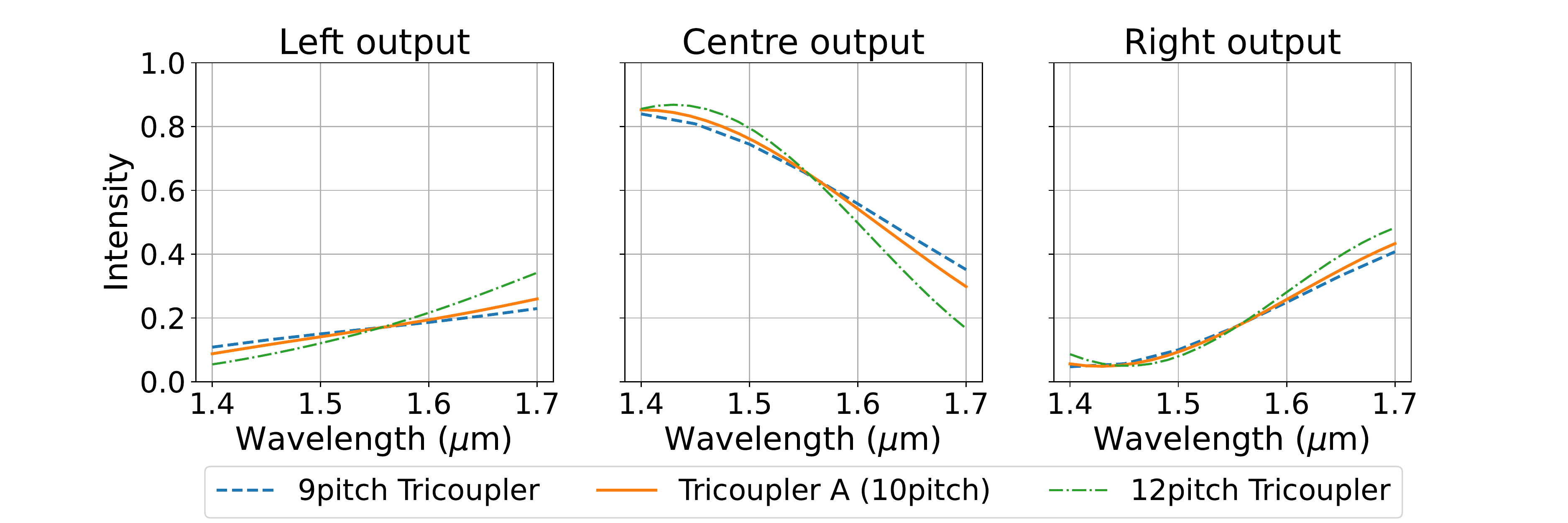}
        \vspace{5pt} 
        \caption{Throughput of an off-axis source located half a fringe away from the phase centre for the bright (left, right) and nulled (centre) outputs and considering three different tricouplers based on the design given in Section~\ref{sec:designs_tricoupler}. While the original, {Tricoupler A}, has a pitch of $10~\mu$m, here we add tricouplers with pitches of 9$~\mu$m and 12$~\mu$m. The interaction lengths of all three tricouplers are designed such that equal splitting occurs at $1.55~\mu$m. Total injected intensity is normalised to 1.}
        \label{fig:planetlight_cplpitch}
    \end{figure}

    The throughputs of simulated planet light for three tricouplers with different pitches were computed, and the result is shown in Fig.~\ref{fig:planetlight_cplpitch}.
    The interaction length for each is optimised such that equal splitting occurs at 1.55$~\mu$m.
    For all three tricouplers, planet light in the centre output has the same throughput: 0.66 at 1.55$~\mu$m.
    Moreover, the intensities in the centre output for {Tricoupler A} vary by 55\%, indicating a strongly chromatic signal across the band.
    While changing the pitch marginally affects the throughput at either end of the waveband, tuning the interaction length to balance left and right outputs at 1.55$~\mu$m has the effect of fixing the throughput to the same intensity in the centre of the band. 
    This limits both the maximum throughput at the central wavelength, and the maximum achievable achromaticity of the signal: the curves in Figure~\ref{fig:planetlight_cplpitch} imply that substantial gains for the planet light signal might require trades against other desired properties of the tricoupler.
    
\begin{figure}[h]
        \centering
        \includegraphics[width=0.9\textwidth]{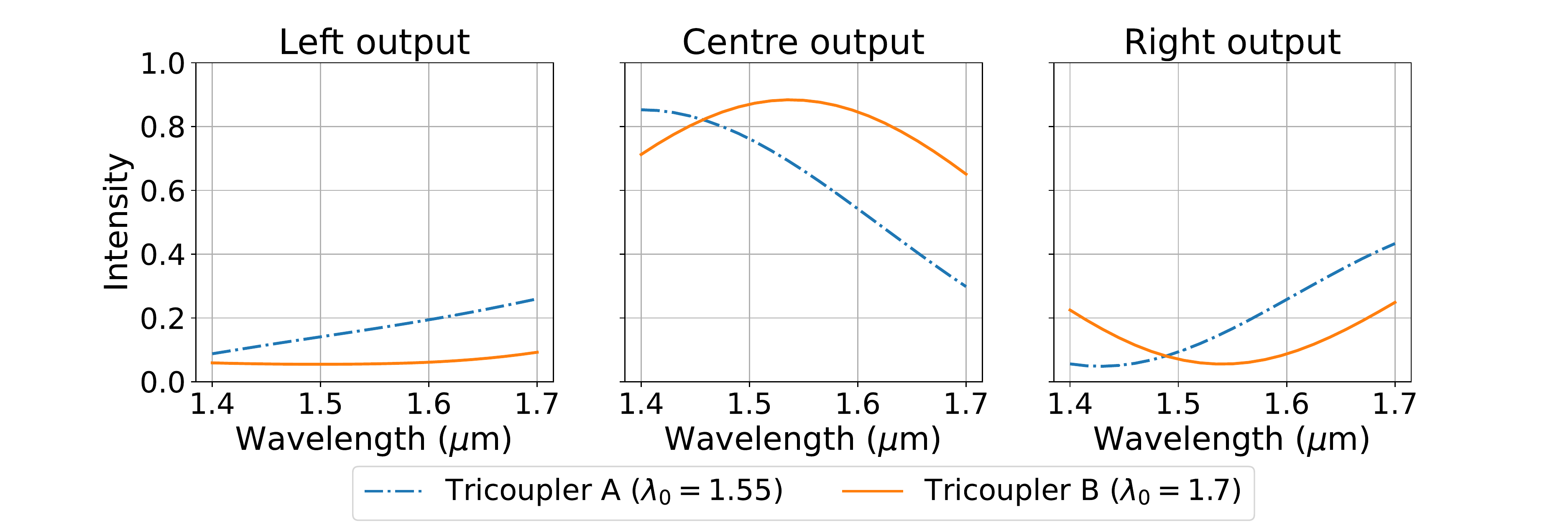}
        \vspace{5pt}
        \caption{Throughput of an off-axis source located half a fringe away from the phase centre in the left, centre and right outputs for two simulated tricouplers: (Blue dashed line) {Tricoupler A} from Section~\ref{sec:designs_tricoupler} designed for 1.55$~\mu$m, and (Orange solid line) {Tricoupler B}, which is tailored for equal splitting at $1.7~\mu$m. Total injected intensity is normalised to 1.}
        \label{fig:planetlight_splitting}
    \end{figure}

    A higher planet light throughput was observed when a tricoupler was tuned such that the equal splitting condition occurred at a different wavelength, through changing the interaction length for a given pitch. 
    Figure \ref{fig:planetlight_splitting} shows a comparison between the planet light output of {Tricoupler A} and a tricoupler whose equal splitting occurs at 1.7$~\mu$m instead of 1.55$~\mu$m, with a pitch of $12~\mu$m and a interaction length of 1122$~\mu$m{, named Tricoupler B}. Here, peak planet light transmission is in the centre of the observing band, with a value at the centre of the band of 0.88. The total planet light throughput integrated across the band is 28\% higher than for the previous  1.55$~\mu$m-tuned tricoupler, and the signal intensity varies by 23\%: more than a factor of two improvement. 
    This latter metric -- the yielding of a flatter spectral response to planet light across the band -- makes it better suited for recovering spectral information for any off-axis stellar companion. 
    
\begin{figure}[h]
        \centering
        \includegraphics[width=0.7\textwidth]{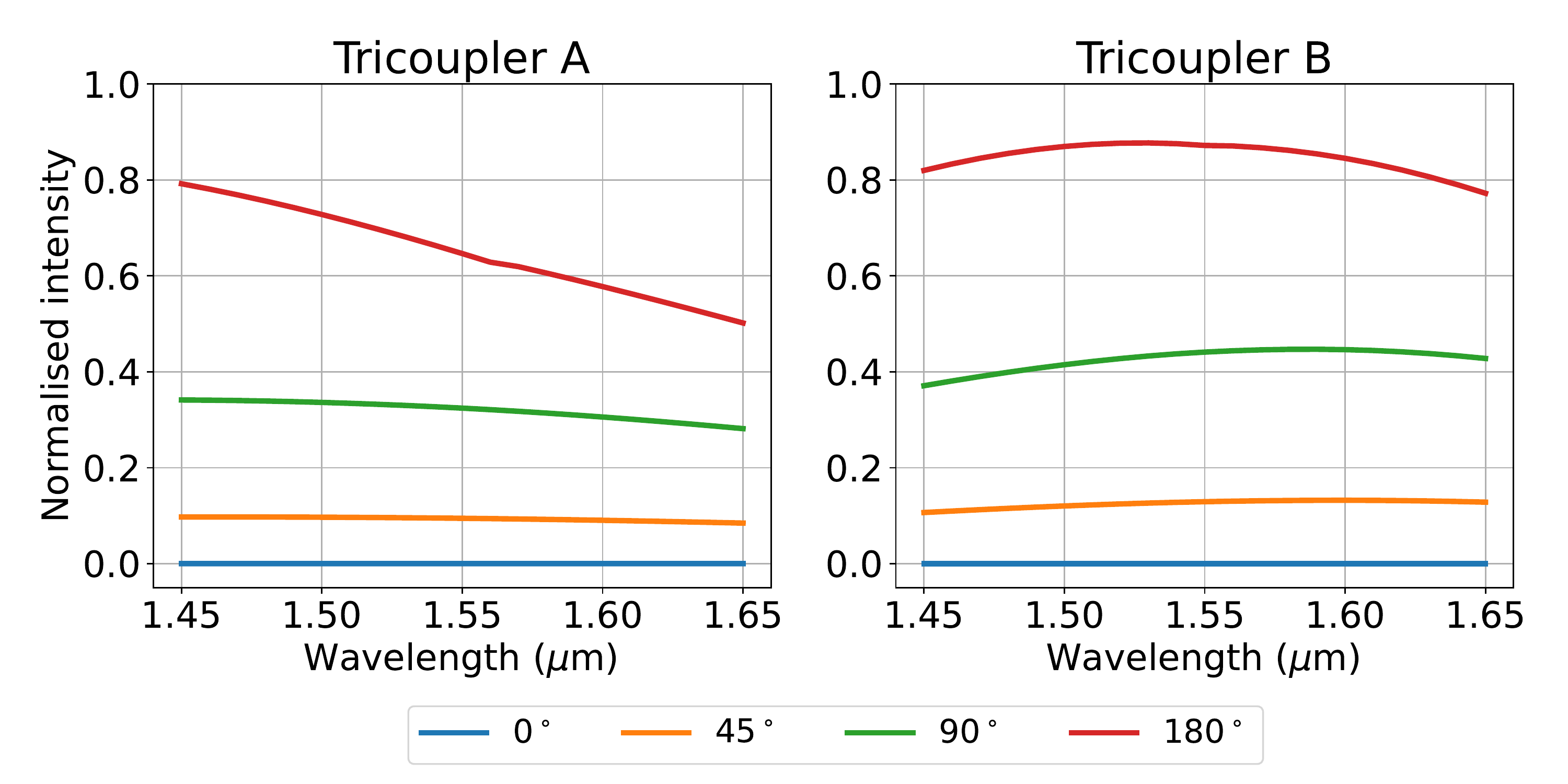}
        \caption{{Throughputs in Tricouplers A (left) and B (right) of light from planetary companions located at different positions relative to their host star: on the same axis as the star ($0^\circ$ phase delay, blue curve), $\frac{1}{8}$ of a fringe away ($45^\circ$ phase delay, orange curve), $\frac{1}{4}$ of a fringe away ($90^\circ$ phase delay, green curve) and half a fringe away ($180^\circ$ phase delay, red curve).}}
        \label{fig:planet_optimisation}
    \end{figure}

    {The design of the tricoupler for increased planet light throughput also holds for different positions of the planet relative to its host star, allowing multiple planet candidates at different separations from their host star to be observed. Figure \mbox{\ref{fig:planet_optimisation}} shows the throughputs of both tricouplers for a planet located at different angular separations from the star: at the first bright fringe, as in Figure \mbox{\ref{fig:planetlight_splitting}}, corresponding to a 180$^\circ$ geometrical phase delay in the tricoupler; halfway between (ie. located a quarter of a fringe away), corresponding to a 90$^\circ$ delay; an eighth of a fringe away (45$^\circ$), and in the plane of the star (0$^\circ$). The closer the planet is to its host star, the less light is detected in the tricoupler. The planet located on the same axis as its star, corresponding to the blue dot in Figure \mbox{\ref{fig:null_modulation}}, is completely nulled along with the starlight. 
    For planets located on the other side of the star (to the left of the star in Figure \mbox{\ref{fig:null_modulation}}), the sign of the incoming phase delay is flipped, so the curves in Figure \mbox{\ref{fig:planet_optimisation}} are flipped horizontally. 
    Hence, this design of the tricoupler allows for planet light throughput to be maximised compared with Tricoupler A, regardless of the location of the planet relative to its host star.}

    \begin{figure}[h]
        \centering
        \includegraphics[width=0.9\textwidth]{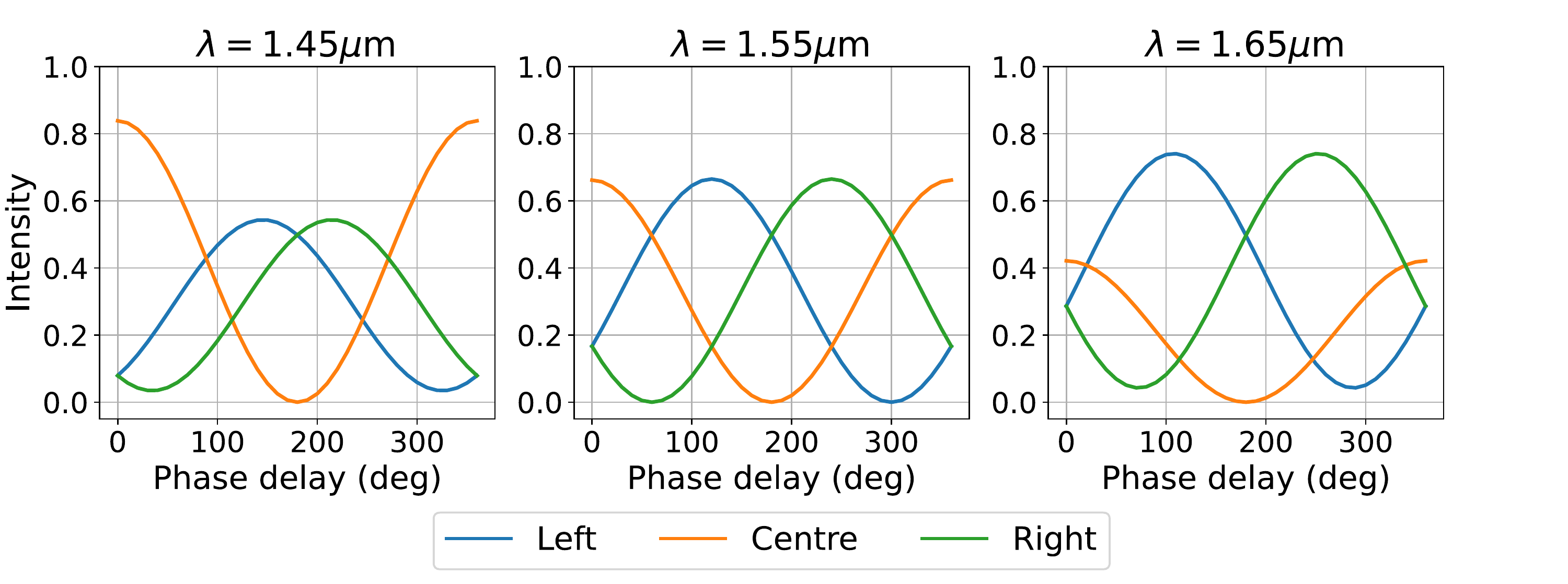}
        \caption{Output intensities of all three channels {of Tricoupler A }at three selected wavelengths: 1.45$~\mu$m, 1.55$~\mu$m and 1.65$~\mu$m.}
        \label{fig:tric_a_sel_wls}
    \end{figure}

    \begin{figure}[h]
        \centering
        \includegraphics[width=0.9\textwidth]{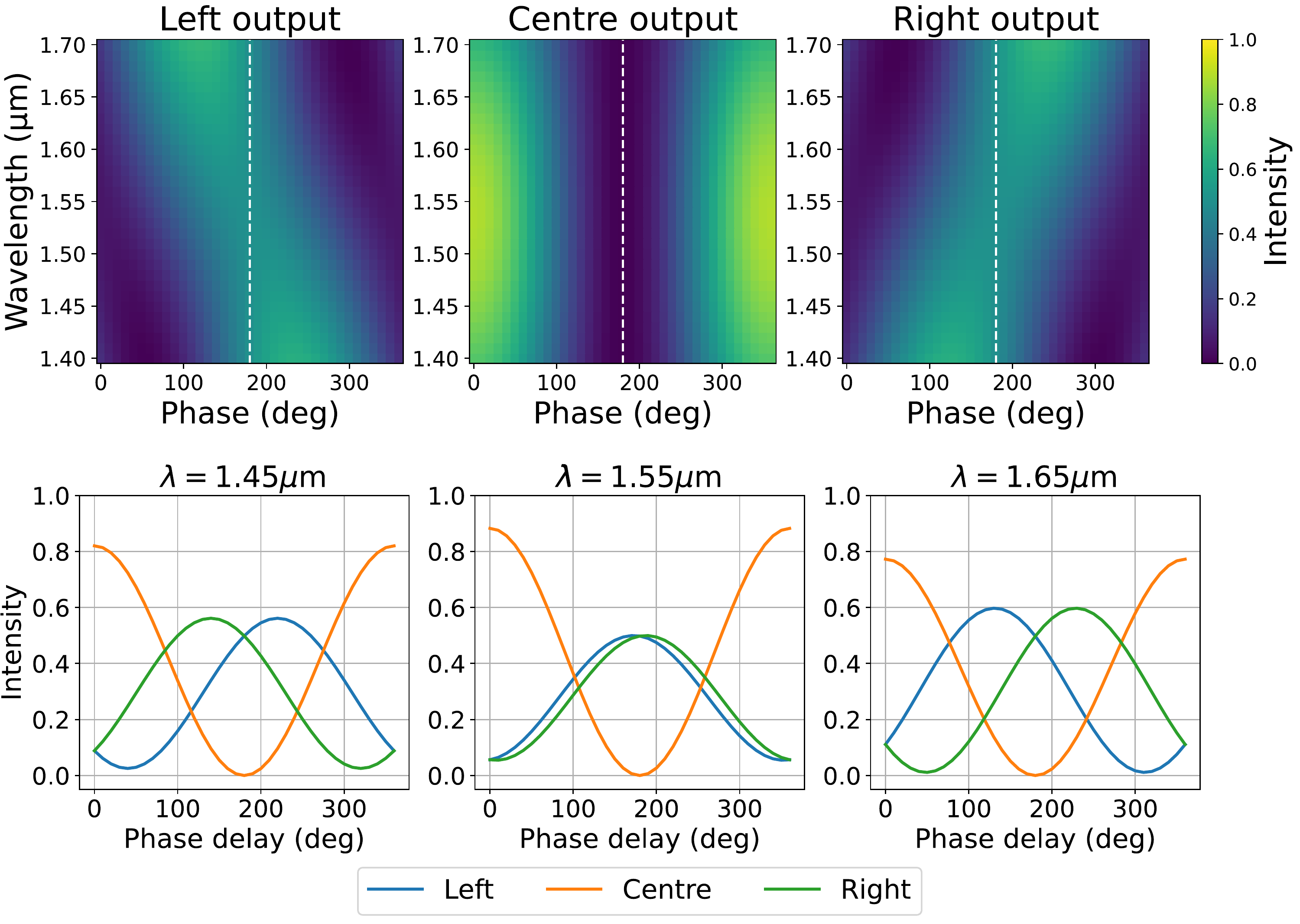}
        \caption{Top: Map of the intensities in the three outputs of {Tricoupler B}, against wavelength and differential phase of incoming beams. Flux at 180$^\circ$ differential phase is shown by the white dashed line. Bottom: Output intensities of all three channels at three selected wavelengths: 1.45$~\mu$m, 1.55$~\mu$m and 1.65$~\mu$m.}
        \label{fig:12-1122_combined_cmap_sel_wls}
    \end{figure}

    However, {the useful extra throughput of Tricoupler B} and flattening of the spectral response for an off-axis source was found to come at the price of losses in the sensitivity of the phase measurement, ultimately impacting the fringe tracking capability of the tricoupler. 
    The output intensities for three selected wavelengths at each end and in the centre of the waveband, 1.45$~\mu$m, 1.55$~\mu$m and 1.65$~\mu$m, are shown in Figure \ref{fig:tric_a_sel_wls}.  
    In an ideal case, the left and right outputs (blue and green) will cross when their gradients are the steepest, meaning that the differential output flux linearly and significantly changes with respect to the differential phase.
    The left and right outputs of {Tricoupler A} cross when almost at their steepest, although less so at shorter wavelengths (Fig.~\ref{fig:tric_a_sel_wls}).
    
    Figure \ref{fig:12-1122_combined_cmap_sel_wls} shows the equivalent output intensities for {Tricoupler B}, where equal splitting is designed to occur at 1.7$~\mu$m and planet light throughput is maximised. 
    The differential flux between left and right is overall shallower than {for Tricoupler A }when the differential phase is around 180$^\circ$ and becomes nonexistent at the centre of the spectral bandwidth.
    Thus, phase metrology needed to track the fringes and therefore maintain a stable null has been compromised compared to {Tricoupler A}.
    This sacrifice has yielded gains as discussed above for the recovery of the planet light signal.
    For any given nulling interferometry experiment, tuning between these various tradeoffs must be informed by practical considerations of the desired science outcomes, the nature of the targets under study, and the expected performance metrics of the wider instrumental system.

\section{Designs for an achromatic phase shifter}\label{sec:designs_aps}
    
    Achieving a null fringe in any spectral channel of a beam combiner requires the control of the relative phase of the input beams by way of some adjustable piston term.
    GLINT currently uses a segmented deformable mirror, with each segment matched to a single input waveguide (in effect, an in-air delay line) to find and maintain this destructive interference null condition.
    Such delay lines are intrinsically chromatic unless they are operated at an optical path difference of zero.
    On the other hand, the fundamentals of our design, as discussed above, requires injected beams to be in anti-phase: a path difference of $180^\circ$.
    No simple delay system can polychromatically achieve this non-zero phase relationship.
    
    We propose to replace the air-delay lines with photonic achromatic phase shifters (APS hereafter) inscribed inside the chip upstream of each tricoupler.
    In-guide delays can be effected by adjusting the effective refractive index, for example by changing the physical properties of the material or by changing the diameter of the waveguide.
    These alterations produce a chromatic effect, as varying delay functions are imposed across the band.
    A phase shifter, altering the differential phase between a pair of waveguides, can be engineered by imposing different changes to one guide compared to the other.
    With careful optimisation, the functional form of these chromatic delays can be balanced to yield a net overall achromatic delay.
    This method was successfully used by the European Southern Observatory's GRAVITY's ABCD beam combiner\cite{gravity2017}, with the design and calculations required extensively described in Ref.~\citenum{labeye2008}.

\begin{figure}[h]
        \centering
        \includegraphics[width=0.7\textwidth]{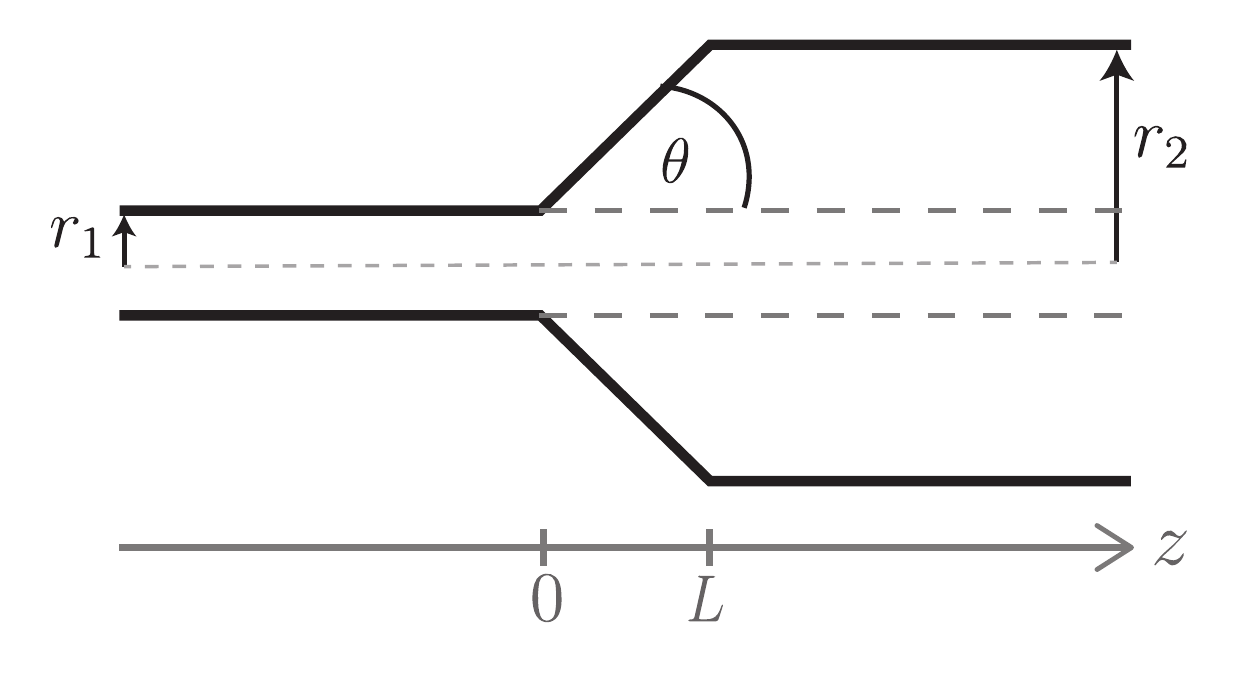}
        \caption{{Schematic of a waveguide with two different radii, $r_1$ and $r_2$, with a tapered section transitioning between these. The angle of the taper is given by $\theta$.}}
        \label{fig:taper_demo}
    \end{figure}

\begin{figure}[h]
        \centering
        \includegraphics[width=0.7\textwidth]{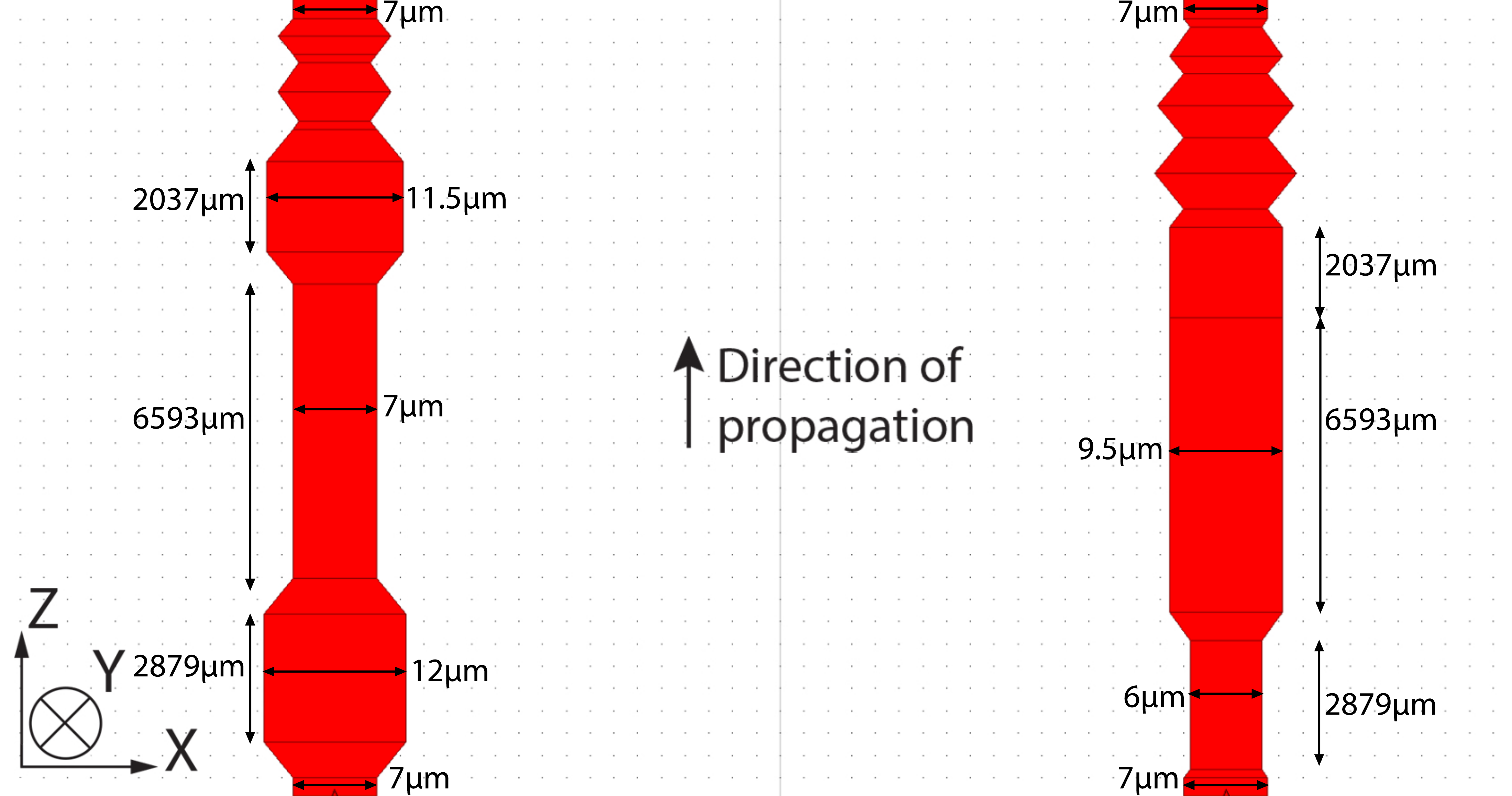}
        \caption{Schematic of the proposed APS. Light is injected at the bottom with the direction of propagation upward. 
        Final sections uppermost in each waveguide contain tapers corresponding to those needed by the other guide, replicated to balance chromatic effects.}
        \label{fig:aps}
    \end{figure}

    Following the method used in Ref.~\citenum{labeye2008}, each waveguide is fabricated with three tailored sections where diameter and length are tuned for refractive index changes.
    The target outcome for the pair of waveguides is an achromatic phase shift across the entire $H$ band, with our best solution illustrated in (Fig.~\ref{fig:aps}).
    The tapered (conical) transitions between sections were added to minimise the loss of light that would otherwise occur at step-function changes in the guide structure.
    The criterion for the minimum taper angle\cite{labeye2008}, $\theta$, for light to pass adiabatically between segments is

        \begin{equation}\label{eqn:adiabacity}
            \theta(z) < \frac{r(z)}{\lambda_{\text{\textbf{max}}}}\De n,
        \end{equation}
    where $r$ is the radius of the waveguide, $\lambda_{\text{min}}$ the smallest wavelength in the band and $\De n$ the difference between background and waveguide refractive index. 
    {Figure \mbox{\ref{fig:taper_demo}} shows a schematic of a waveguide, where a tapered section transitions between two different radii. From this, the following relationship can also be found:}
    \begin{equation}
        \tan(\theta(z)) = \frac{r_2 - r_1}{L},
    \end{equation}
    {with $r_2 > r_1$ the radii of the two waveguide sections.}
    
    Using this relationship, a threshold minimum length for each taper section{ is given as:}
    
        \begin{equation}\label{eqn:adiabacity_length}
            L > \frac{r_2 - r_1}{\tan\left( \frac{r_1}{\lambda_{\text{\textbf{max}}}\De n} \right)}.
        \end{equation}
        
    As $L$ depends on the waveguide radii, the minimum length of each taper depends on the widths of the two waveguide sections on either side of it. 
    While these taper sections do mitigate losses, they also have the unfortunate effect of introducing further phase chromaticity\cite{labeye2008}.
    To combat this, each taper segment was implemented on both guides so that both wavefronts undergo identical taper-induced chromatic phase effects.
    The way this is implemented is shown in Figure \ref{fig:aps}: for each guide taper sections belonging to the other guide were added onto the end.
    
\begin{figure}[h] 
        \centering
        \includegraphics[width=0.7\textwidth]{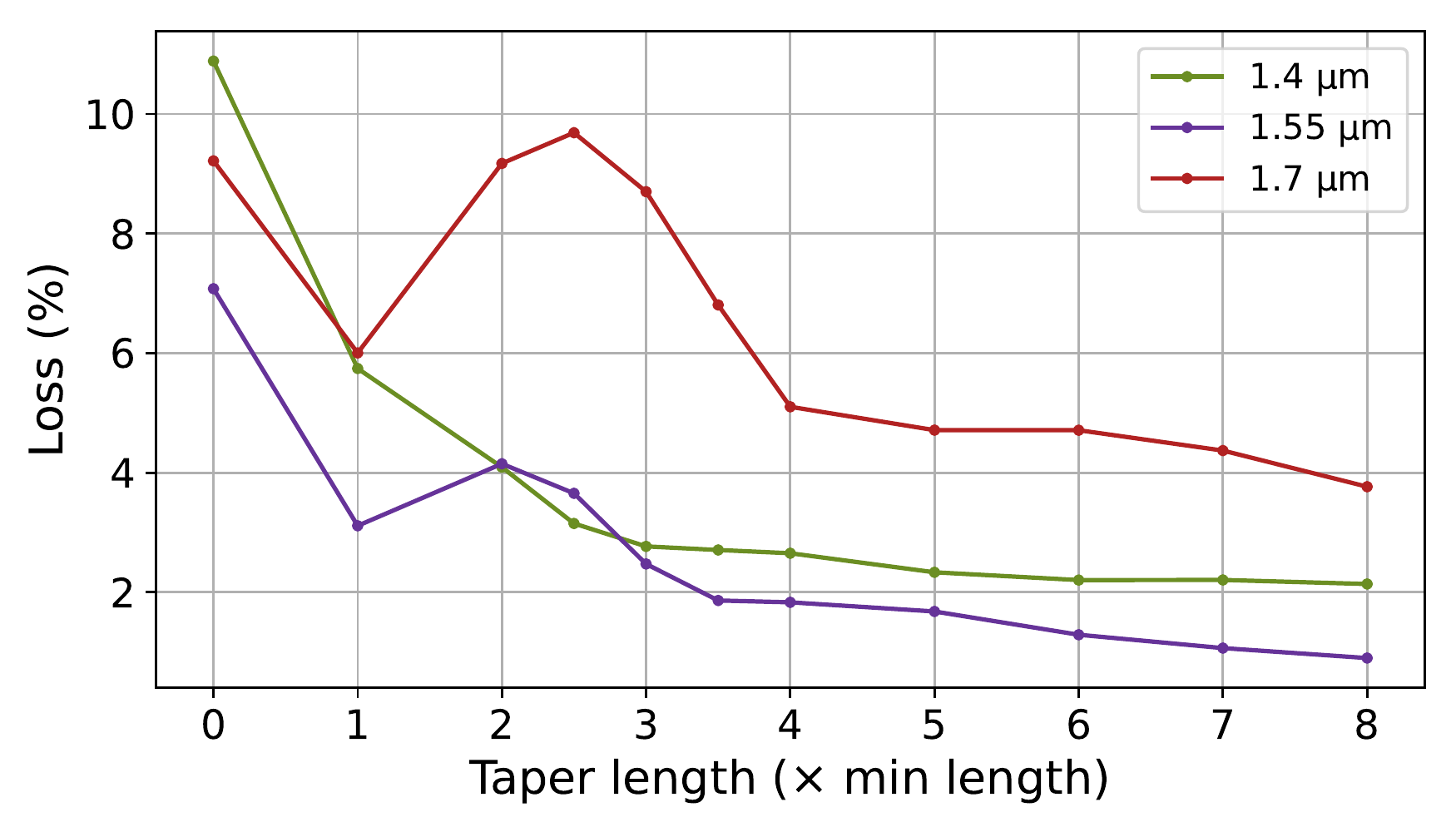}
        \caption{Total loss of light due to propagation in the APS as a function of lengths of taper sections in multiples of the minimum length, as defined by Equation \ref{eqn:adiabacity_length}. This loss is plotted for the central wavelength ($1.55~\mu$m) as well as the upper and lower-bound wavelengths, $1.4~\mu$m and $1.7~\mu$m.}
        \label{fig:loss_vs_aps_length}
    \end{figure}

    To optimise device throughput, light loss over the phase shifter was plotted while varying the total length as computed by \texttt{RSoft} for the central wavelength, as well as the upper- and lower-bound wavelengths of the band, and is shown in Figure \ref{fig:loss_vs_aps_length}.
    Beyond a length of roughly $2.5$ times the minimum taper length, the loss in the phase shifter decreases with increasing taper length (corresponding to increasing overall APS length) for all wavelengths, however it plateaus at a value with diminishing returns past four times the minimum length. Hence, the optimum taper length adopted was a scaling of four times the minimum from Equation~\ref{eqn:adiabacity_length}.
    The total length of the APS including these taper lengths is $18$\,mm.
    {This design is a first step toward a functional APS made with ULI, hence some parameters may need to change based on experimental constraints while keeping an achromatic phase shift of 180$^\circ$ and for shortening the device.}
    
\begin{figure}[h]
        \centering
        \includegraphics[width=0.7\textwidth]{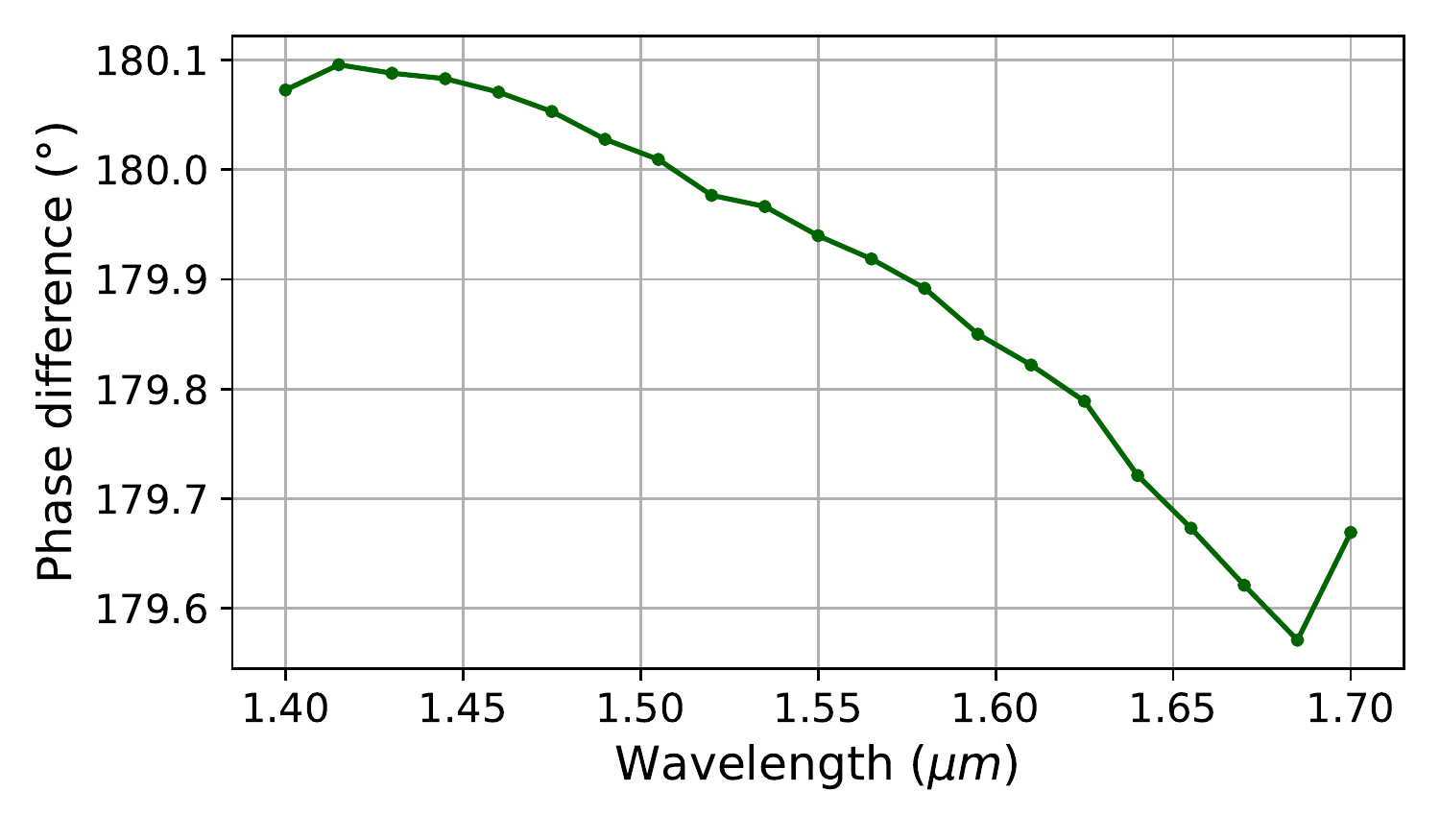}
        \caption{{Differential phase between wavefronts on the output side of the APS device with taper lengths of four times the minimum length, as a function of wavelength.}}
        \label{fig:aps_chromaticity}
    \end{figure}

    {Simulations of the performance of the full, chromatically-symmetrised device with the lengths of tapered sections at four times the minimum length were performed in \texttt{RSoft}.}
    The key performance indicator, the phase difference at the output over the waveband of $1.4-1.7~\mu$m, was found to vary by less than 0.6$^\circ$, as illustrated in Fig.~\ref{fig:aps_chromaticity}. 
    This allows for an effectively achromatic phase shift under the expected working conditions.

\section{Integrating components into a fully-functioning device}\label{sec:device}
    
\begin{figure}[h]
        \centering
        \includegraphics[width=0.6\textwidth]{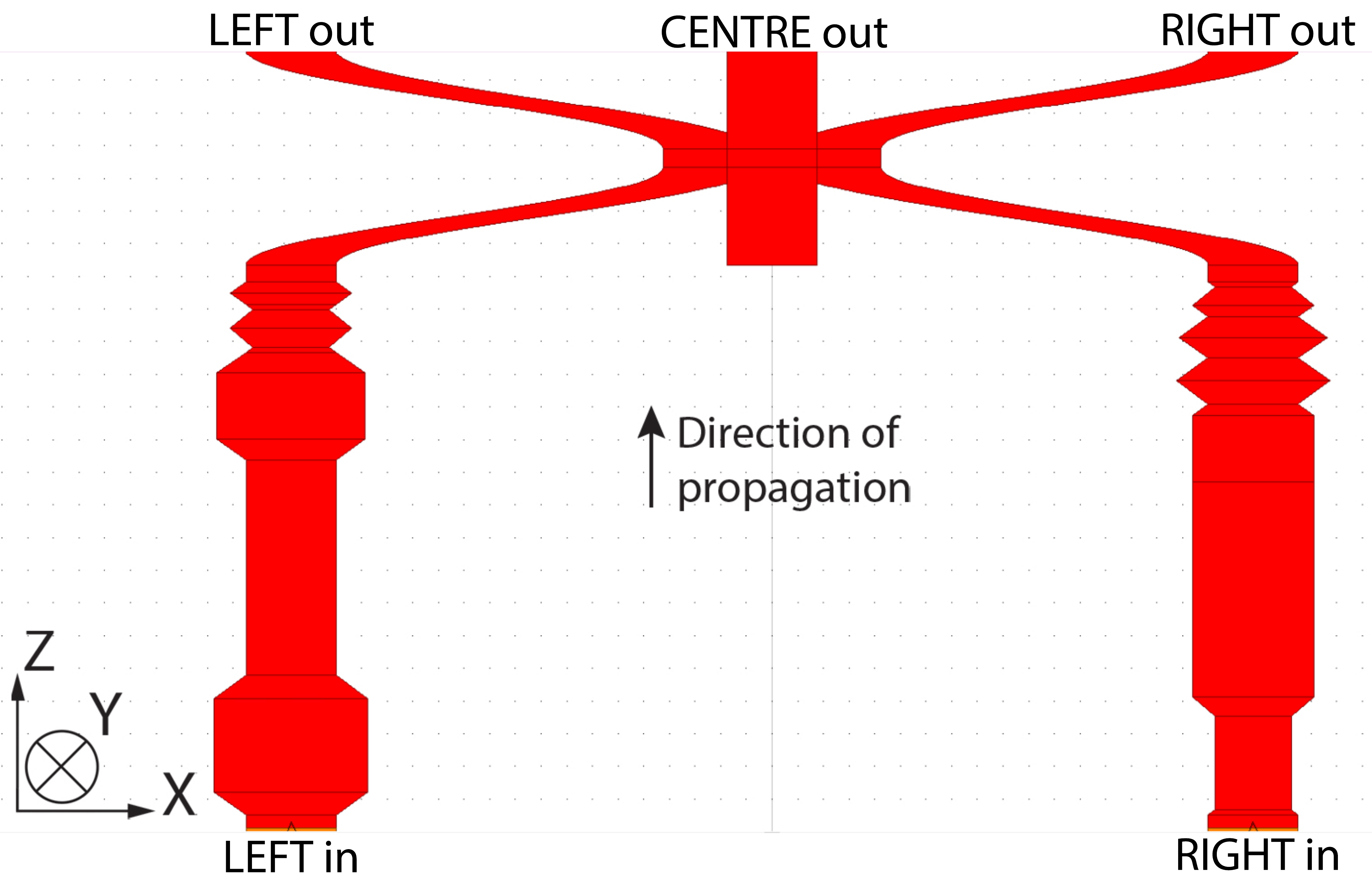}
        \caption{Schematic of the proposed combined APS (lower regions of the diagram) and {Tricoupler A} (upper region).}
        \label{fig:combined}
    \end{figure}

    Since it is designed to be fabricated using the same ULI process as required by the 3D tricoupler design proposed in Section~\ref{sec:designs_tricoupler}, the APS design above can be incorporated into the same chip as the tricoupler.
    A layout of both components sequentially integrated into a single device is shown in Figure \ref{fig:combined}.
    Two beams of starlight are fed into the left and right inputs at the bottom of the diagram, firstly undergoing the $180^\circ$ achromatic phase shift in the APS section, and are subsequently combined in the tricoupler section.
    Finally, the left, central and right outputs exit the chip at the top, where they would progress through to a sensor.
    
    \begin{figure}[h]
        \centering
        \includegraphics[width=0.7\textwidth]{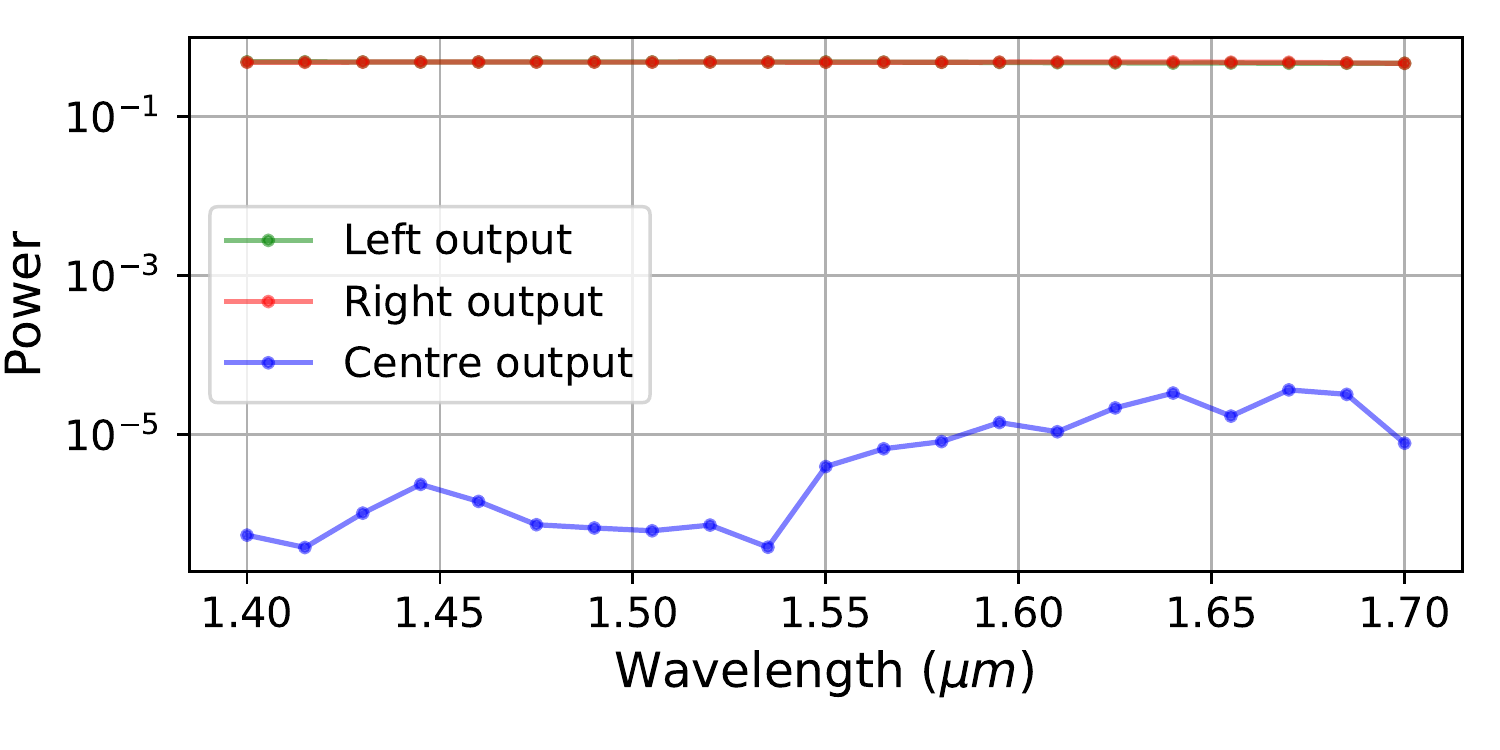}
        \caption{{Power in each output for the combined APS and tricoupler device, as a function of wavelength. Total power is shown on a logarithmic scale and normalised to 1.}}
        \label{fig:combined_pow_vs_wl}
    \end{figure}

    {The full optimised architecture of an integrated device consisting of the APS and Tricoupler A (as shown in Figure~\mbox{\ref{fig:combined}}) was then numerically tested in the  \texttt{RSoft} package.}
    Results are plotted in Fig.~\ref{fig:combined_pow_vs_wl}, where the simulation has been framed to illustrate the ideal working condition for the overall device: injection by way of twin beams of equal intensity and phase.
    Performance as shown in Fig.~\ref{fig:combined_pow_vs_wl} can be seen to be close to ideal: very little light is found in the central (null) channel, with the maximum output intensity in this channel at $10^{-4}$ with respect to total input flux. Output signals are nearly equally split over the left and right guides. 
    Encouragingly, these design outcomes appear to be nearly achromatic with the same splitting persisting across the operational waveband. 
    
    The total length of the combined device is $24.6$\,mm. In \texttt{RSoft} simulations, the device has a total throughput of $94.5\%$, which takes into account loss due to waveguide bending, but not propagation through the material. 
    Using an upper bound for waveguide propagation losses of $0.14\pm0.04$dB/cm\cite{norris2020first}, the total throughput including bend and propagation losses is calculated to be $87.3\%$. 
    
    Once parameters to fabricate a working model have been tuned, a compact laser-written device implementing the combined phase shifter/tricoupler in Figure \ref{fig:combined} can be replicated multiple times within a single chip.
    This enables nulling interferometry to be performed on several baselines at once. 
    As the $180^\circ$ phase shift is introduced directly before the tricoupler -- rather than outside the photonic chip as in the current design of GLINT -- all baselines can be nulled simultaneously.

\section{Conclusion}\label{sec:concl}

    With further development, nullers like the GLINT instrument have the potential to reach the scales of contrast and angular resolution needed for the direct imaging of exoplanets. 
    The use of a photonic tricoupler on the nulling chip overcomes several significant challenges in reaching this contrast, including the chromaticity inherent in traditional directional couplers and real-time metrology to combat phase fluctuations. 
    A new design of the tricoupler has been proposed, which incorporates full equilateral symmetry for simplicity in interactions and flexibility in waveguide injection. 
    The benefits of optimising for either the off-axis light throughput or the fringe tracking capability have also been explored, with the trade space between these highlighted and clarified for optimisation as desired specific to the details of any science application. 

    To implement the achromatic $180^\circ$ phase shift required for nulling, a photonic phase shifter has been designed, making use of tapered waveguides of differing diameters to impose tailored variations in the effective index. 
    This method allows the phase shifter to be integrated into a single chip, directly upstream of the tricoupler. 
    The phase shifter achieves a $180^\circ$ phase shift with a variation of $0.6^\circ$ across the waveband when simulated.
    {Work remains to be done to shorten the length of the phase shifter in order to ease the scalability and improve the throughput.}
    The phase shifter and tricoupler presented here, integrated into a single component, can readily be replicated several times on a single photonic chip for use in simultaneous multi-baseline nulling.

\subsection*{Disclosures}
The authors declare no conflicts of interest.

\subsection* {Code, Data, and Materials Availability} 
Data underlying the results presented in this paper are not publicly available at this time but may be obtained from the authors upon reasonable request.


\bibliography{report}   

\begin{thebibliography}{10}

\bibitem{bracewell1978detecting}
R.~N. Bracewell, ``Detecting nonsolar planets by spinning infrared
  interferometer,'' {\em Nature} {\bf 274}(5673), 780--781  (1978).

\bibitem{norris2020first}
B.~R. Norris, N.~Cvetojevic, T.~Lagadec, {\em et~al.}, ``First on-sky
  demonstration of an integrated-photonic nulling interferometer: the glint
  instrument,'' {\em Monthly Notices of the Royal Astronomical Society} {\bf
  491}(3), 4180--4193  (2020).

\bibitem{Marois2008}
C.~{Marois}, B.~{Macintosh}, T.~{Barman}, {\em et~al.}, ``{Direct Imaging of
  Multiple Planets Orbiting the Star HR 8799},'' {\em Science} {\bf 322}, 1348
  (2008).

\bibitem{Schworer2015}
G.~{Schworer} and P.~G. {Tuthill}, ``{Predicting exoplanet observability in
  time, contrast, separation, and polarization, in scattered light},'' {\em
  Astron. Astrophys} {\bf 578}, A59  (2015).

\bibitem{colavita2009keck}
M.~Colavita, E.~Serabyn, R.~Millan-Gabet, {\em et~al.}, ``Keck interferometer
  nuller data reduction and on-sky performance,'' {\em Publications of the
  Astronomical Society of the Pacific} {\bf 121}(884), 1120  (2009).

\bibitem{defrere2016nulling}
D.~Defrere, P.~Hinz, B.~Mennesson, {\em et~al.}, ``Nulling data reduction and
  on-sky performance of the large binocular telescope interferometer,'' {\em
  The Astrophysical Journal} {\bf 824}(2), 66  (2016).

\bibitem{mennesson2011high}
B.~Mennesson, C.~Hanot, E.~Serabyn, {\em et~al.}, ``High-contrast stellar
  observations within the diffraction limit at the palomar hale telescope,''
  {\em The Astrophysical Journal} {\bf 743}(2), 178  (2011).

\bibitem{martinod2021scalable}
M.-A. Martinod, B.~Norris, P.~Tuthill, {\em et~al.}, ``Scalable photonic-based
  nulling interferometry with the dispersed multi-baseline glint instrument,''
  {\em Nature communications} {\bf 12}(1), 1--11  (2021).

\bibitem{serabyn2000nulling}
E.~Serabyn, ``Nulling interferometry: symmetry requirements and experimental
  results,'' in {\em Interferometry in optical astronomy},   {\bf 4006},
  328--339, International Society for Optics and Photonics  (2000).

\bibitem{martinod2021achromatic}
M.-A. Martinod, P.~Tuthill, S.~Gross, {\em et~al.}, ``Achromatic photonic
  tricouplers for application in nulling interferometry,'' {\em Applied Optics}
  {\bf 60}(19), D100--D107  (2021).

\bibitem{labeye2004}
P.~R. {Labeye}, J.-P. {Berger}, M.~{Salhi}, {\em et~al.}, ``{Integrated optics
  components in silica on silicon technology for stellar interferometry},'' in
  {\em New Frontiers in Stellar Interferometry},  W.~A. {Traub}, Ed., {\em
  Proc. SPIE} {\bf 5491}, 667  (2004).

\bibitem{weber2004}
V.~{Weber}, M.~{Barillot}, P.~{Haguenauer}, {\em et~al.}, ``{Nulling
  interferometer based on an integrated optics combiner},'' in {\em New
  Frontiers in Stellar Interferometry},  W.~A. {Traub}, Ed., {\em Society of
  Photo-Optical Instrumentation Engineers (SPIE) Conference Series} {\bf 5491},
  842  (2004).

\bibitem{vance1994design}
R.~{Vance} and J.~{Love}, ``Design procedures for passive planar coupled
  waveguide devices,'' {\em IEE Proceedings - Optoelectronics} {\bf 141},
  231--241(10)  (1994).

\bibitem{hsiao2010}
H.-K. {Hsiao}, K.~A. {Winick}, and J.~D. {Monnier}, ``{Midinfrared broadband
  achromatic astronomical beam combiner for nulling interferometry},'' {\em
  Appl. Opt.} {\bf 49}, 6675  (2010).

\bibitem{birks1992effect}
T.~Birks, ``Effect of twist in 3$\times$ 3 fused tapered couplers,'' {\em
  Applied optics} {\bf 31}(16), 3004--3014  (1992).

\bibitem{nolte2003femtosecond}
S.~Nolte, M.~Will, J.~Burghoff, {\em et~al.}, ``Femtosecond waveguide writing:
  a new avenue to three-dimensional integrated optics,'' {\em Applied Physics
  A} {\bf 77}(1), 109--111  (2003).

\bibitem{gattass2008femtosecond}
R.~R. Gattass and E.~Mazur, ``Femtosecond laser micromachining in transparent
  materials,'' {\em Nature photonics} {\bf 2}(4), 219--225  (2008).

\bibitem{arriola2013low}
A.~Arriola, S.~Gross, N.~Jovanovic, {\em et~al.}, ``Low bend loss waveguides
  enable compact, efficient 3d photonic chips,'' {\em Optics express} {\bf
  21}(3), 2978--2986  (2013).

\bibitem{gross2015ultrafast}
S.~Gross and M.~Withford, ``Ultrafast-laser-inscribed 3d integrated photonics:
  challenges and emerging applications,'' {\em Nanophotonics} {\bf 4}(3),
  332--352  (2015).

\bibitem{Hansen2020}
J.~T. {Hansen}, M.~J. {Ireland}, T.~{Travouillon}, {\em et~al.}, ``{Linear
  formation-flying astronomical interferometry in low-Earth orbit: a
  feasibility study},'' in {\em Society of Photo-Optical Instrumentation
  Engineers (SPIE) Conference Series},  {\em Society of Photo-Optical
  Instrumentation Engineers (SPIE) Conference Series} {\bf 11443}, 1144366
  (2020).

\bibitem{hansen2022interferometric}
J.~T. Hansen, M.~J. Ireland, A.~Ross-Adams, {\em et~al.}, ``Interferometric
  beam combination with a triangular tricoupler photonic chip,'' {\em Journal
  of Astronomical Telescopes, Instruments, and Systems} {\bf 8}(2), 025002
  (2022).

\bibitem{fang1996interferometric}
X.~Fang, R.~O. Claus, and G.~Indebetouw, ``Interferometric model for phase
  analysis in fiber couplers,'' {\em Applied optics} {\bf 35}(22), 4510--4515
  (1996).

\bibitem{gravity2017}
{Gravity Collaboration}, R.~{Abuter}, M.~{Accardo}, {\em et~al.}, ``{First
  light for GRAVITY: Phase referencing optical interferometry for the Very
  Large Telescope Interferometer},'' {\em \aap} {\bf 602}, A94  (2017).

\bibitem{labeye2008}
P.~Labeye, {\em {Composants optiques int{\'e}gr{\'e}s pour
  l'Interf{\'e}rom{\'e}trie astronomique}}.
\newblock Theses, {Institut National Polytechnique de Grenoble - INPG}  (2008).

\end{thebibliography}
\bibliographystyle{spiejour}   


\end{spacing}
\end{document}